**$T_1$, $T_2$, and complex permittivities of hydrogels, paramagnetic salt solutions, and oils at 0.35, 1.5, and 3 Tesla**

H. Michael Gach[a)]

*Departments of Radiation Oncology, Radiology, and Biomedical Engineering, Washington University in St. Louis, St. Louis, Missouri 63110, Email: hmgach@gmail.com*

Kerri Prinos[b)]

*Department of Electrical and Systems Engineering, Washington University in St. Louis, St. Louis, Missouri 63110, Email: kerriprinos@icloud.com,*

Hannah Sechaga[c)]

*Department of Radiation Oncology, Washington University School of Medicine, St. Louis, Missouri 63110, Email: hannahbirhanu38@gmail.com*

[a)] Author to whom correspondence should be addressed.

[b)] Now at Horizon3.ai

[c)]Now at Baylor Scott & White Health, Dallas, Texas 75204

Author Contributions: H MIchael Gach: Research design, research funding, pulse sequence development, data acquisition and analysis, manuscript composition and editing. Kerri Prinos: Data acquisition and analysis, manuscript composition. Hannah Sechaga: Data acquisition and analysis.

**ACKNOWLEDGMENTS**

This research was funded in part under National Institutes of Health (NIH) National Heart, Lung, and Blood Institute (NHLBI) grant R01 HL148210. This research was conducted under a Master Research Agreement (MRA) between Washington University in St. Louis and ViewRay. The research was supported by the Washington University in St. Louis Departments of Radiation Oncology and Radiology. The studies presented in this work were conducted in part using the resources of the MIR Center for Clinical Imaging Research located at the Washington University Medical Center.

**DISCLOSURE OF CONFLICTS OF INTEREST**

The authors have no relevant conflicts of interest to disclose.

**DATA AVAILABILITY**: Data may be provided upon request.

**HIGHLIGHTS:** Relaxation times and complex permittivities for three types of materials (hydrogels, paramagnetic salt solutions, and oils) commonly used in MRI phantoms are measured at three magnetic field strengths (0.35, 1.5, and 3 T), interpolated for 0.55 T, and extrapolated for 7 T. Assumptions that $T_1$ rises, or $T_1$ relaxivity ($r_1$) decreases, with increasing field strength was not true for some of our samples under test.

**Abstract:**

**Approach:** The $T_1$, $T_2$, and complex permittivities of candidate MRI phantom constituents (hydrogels, paramagnetic electrolytes, and oils) were measured at different magnetic field strengths and concentrations.

**Data:** $T_1$ decreased with increasing concentration except for PEG. The $T_1$ and $T_2$ relaxivities of polyethylene glycol (PEG), polyvinylpyrrolidone (PVP), sodium polyacrylate, and deionized water did not significantly vary with field strength. The $T_2$ relaxivities of gelatin and sodium alginate did not significantly vary with field strength. All of the hydrogels and paramagnetic electrolyte solutions had high static dielectrics (e.g., $\varepsilon_s$~70) similar to water. The static electrical conductivities of Miller's LB agar, gelatin, PVA, sodium alginate, sodium polyacrylate, xanthan gum, $CuSO_4$, $NiCl_2$, and $Mn(NO_3)_2$ rose with concentration. However, the conductivities of PEG and PVP did not rise with concentration.

**Conclusions:** Sodium polyacrylate, PVP, and PEG hydrogels are challenging for generating consistent phantoms. Sodium alginate, Miller's LB agar, and xanthan gum were good hydrogel candidates. $Mn(NO_3)_2$ had the highest relaxivities of the tested samples. All three oils (canola, castor, and grapeseed) are good candidates for low dielectric ($\varepsilon_s$<10) phantoms for high field applications) but with low conductivities.

## 1. INTRODUCTION

Phantoms are an essential tool for MRI quality assurance (QA). A large variety of commercial MRI phantoms are available. However, commercial phantoms can be expensive especially for MRI facilities with limited funding. Home-built phantoms may be satisfactory for many applications and their materials may be cheap.

The design of a phantom requires knowledge about the material relaxation times and permittivities. Phantom materials include water-based liquids doped with paramagnetic electrolytes, oils, and gels. Material properties like relaxation times and permittivities are often hard to find at clinical Larmor frequencies.

The purpose of this study was to measure and compare the $T_1$, $T_2$, and complex permittivities of hydrogels, paramagnetic electrolytes, and oils commonly used in MRI phantoms. Many of the relaxation publications occurred prior to the era of 1.5 and 3 T clinical MRI. Our measurements cover nominal clinical field strengths (1.5 and 3 T) and 0.35 T used in MRI-guided radiation therapy (MRgRT). Other desirable clinical field strengths include 7 T (high field) and 0.55 T (low field). Therefore, extrapolations (7 T) and interpolations (0.55 T) of $R_1$ and $r_1$ were performed and are included below.

## 2. METHODS

Relaxation times and relaxivities were measured at 21°C for three field strengths: 0.35 T (ViewRay MRIdian MRI-Linac, VB19), 1.5 T (Philips Ingenia MRI, V5.7), and 3 T (Siemens mMR PET-MRI, VE11P).

$T_1$ was measured using an inversion recovery (IR) echo planar imaging (EPI) sequence with 26-30 variable inversion times (TIs) ranging from 21 ms to 5 s. $T_2$ was measured using a Carr–Purcell–Meiboom–Gill (CPMG) multiple spin-echo sequence with 32 echoes and echo spacing (ΔTE) of 20 ms except for deionized water (ΔTE = 40 ms) and $Mn(NO_3)_2$ samples (ΔTE = 12.2, 9 and 8.8 ms for 0.35, 1.5, and 3 T, respectively). The pulse sequence parameters are listed in Tables 1 and 2. A single slice was acquired to minimize the effects of crosstalk and off-resonance saturation on the results.

$T_1$ and $T_2$ values were fit to their respective three parameter models using a nonlinear fit and Mathematica (Wolfram, V11.1):[1]

$$S(TI) = A \cdot e^{-\frac{TI}{T_1}} + B \tag{1}$$

$$S_{corr}(TE) = C \cdot \left[1 - 2e^{\frac{-\left(TR - \frac{TE}{2}\right)}{T_{1fi}}} + e^{\frac{-TR}{T_{1fit}}}\right] e^{-\frac{TE}{T_2}} + D \tag{2}$$

where TR is the repetition time, TE is the echo time, TI is the inversion time, and *A*, *B*, *C*, *D* are fit variables. $T_{1fit}$ resulted from solving Eq. (1).

$T_1$ values were interpolated (0.55 T) and extrapolated (7 T) $T_1$ from the acquired data using:[2]

$$T_1(B_0) = A \cdot {B_0}^C \tag{3}$$

where *A* and *C* are fit variables and $B_0$ is the field strength. The interpolated and extrapolated $T_1$ values were converted into relaxation rates ($R_1$) and relaxivities ($r_1$).

**Hydrogels:** Hydrogel concentrations were based on prior study ranges, preparation constraints, and our ability to acquire relaxation and permittivity data from the samples. Hydrogels included sodium polyacrylate (Educational Innovations, Water-Loc™, Item No. GB-6A, $19.95/454 g, Supplier: TeacherSource, Concentrations: 0.25-5% w/w)[3], xanthan gum (MP Biomedicals, Stock No. ICN960002180, CAS 11138-66-2, $76.80/100 g, molecular weight (MW): >2,000,000 g/mol, Supplier: Fisher Scientific, Concentrations: 0.5-5% w/w)[4], polyvinyl alcohol (PVA, Aldon Corp SE, Stock No. 470302-040, CAS 9002-89-5, $40.15/500 g, MW: 100,000 g/mol, Supplier: VWR, Concentrations: 0.5-5% w/w)[5], polyethylene glycol (PEG) 8000 (ThermoFisher, Stock No. 043443.36, CAS: 25322-68-3, $58.20/500 g, molecular weight (MW): 62.07 g/mol, Supplier: Fisher Scientific, Concentrations: 1-20 mM)[6], and polyvinylpyrrolidone (PVP, K30 (VWR, Stock No. 97061-820, CAS 9003-39-8, $103.13/500 g, MW: 40,000 g/mol, Supplier: VWR, Concentrations: 10-30% w/w).[7] The gelling agent was combined with deionized water at room temperature and stirred vigorously for a period of time that resulted in homogeneity (i.e., 10 to 180 minutes).[5,8,9] Deionized water (14.3 MΩ-cm) was acquired from a Purelab Pulse deionizer (Elga Labwater, UK).

TABLE 1 $T_1$ inversion recovery pulse sequence parameters

| **Parameter** | **0.35 T** | **1.5 T** | **3 T** |
|---|---|---|---|
| TE | 18 ms | 15 ms | 13 or 20 ms |
| TR | 15 s | 15 s | 15 s |
| Flip angle | 90° | 90° | 90° |
| TI's | 21-5000 ms<br>21,25,30,35,40,50,60,70,<br>80,90,100,120,140,160,<br>180,200,250,300,350,<br>400,450,500,600,750,<br>1000,1500,2000,3000,<br>4000,5000 | 30-5000 ms<br>30,35,40,50,60,70,80,<br>90,100,120,140,160,<br>180,200,250,300,350,<br>400,450,500,600,750,<br>1000,2000,3000,4000,<br>5000 | 21 or 34-5000 ms<br>21/34,40,50,60,70,80,90,<br>100,120,140,160,180,<br>200,250,300,350,400,<br>450,500,600,750,1000,<br>2000,3000,4000,5000 |
| Bandwidth | 2083 Hz/px | 2050 Hz/px | 3030 Hz/px |
| Resolution | 3x3x5 $mm^3$ | 3x3x5 $mm^3$ | 3x3x5 $mm^3$ |

Hydrogels were also prepared by mixing granulated Miller's LB agar (Fisher BioReagents, Lot No. 145037, $223/500 g, Supplier: Fisher Scientific, Concentrations: 0.5-5% w/w)[10] or gelatin powder (Aldon Corp SE, Stock No. 470301-134, CAS 9000-70-8, $35.25/500 g, Supplier: VWR, Concentrations: 0.5-5% w/w)[11] with deionized water at room temperature. Miller's LB agar is a nutrient-rich microbial growth medium which is 32% w/w agar and 27% w/w NaCl. The mixtures were then heated in a microwave oven until they began to boil.[12,13] This sequence ensured even distribution of the gelatin powder, in particular. The mixture was then stirred and placed in the freezer at -20ºC for 20 minutes to allow the gel to set.

**Paramagnetic electrolytes:** The candidate liquids were aqueous solutions of $Mn(NO_3)_2 * 4H_2O$, 98% (Alfa Aesar, Stock No. AAA1852130, CAS 20694-39-7, $53.90/250 g, molecular weight (MW): 251.006 g/mol, 98% purity, Supplier: Fisher Scientific), $NiCl_2 * 6H_2O$, 99.3% (metals basis) (Alfa Aesar, Stock No. AA1468722, CAS

7791-20-0, $114.82/100g, MW: 237.71 g/mol, melting point: 140ºC- $H_2O$, Supplier: Fisher Scientific), and $CuSO_4$ * $5H_2O$ (Zep Root Killer, CAS 7758-99-8, MW: 249.68 g/mol, $13.55/907 g).[4,14,15] Concentrations ranged from 0.5 mM to 20 mM for $NiCl_2$ * $6H_2O$ and $CuSO_4$ * $5H_2O$, and 0.01 to 5 mM for $Mn(NO_3)_2$.

**Oils:** The oils tested were food-grade canola oil (365 by Whole Foods Market, Organic Expeller Pressed Canola Oil, Lot No. 1160443, $3.99/473 mL, Supplier: Whole Foods Market, molecular weight (MW): 891.45 g/mol, density 0.915 g/cm$^3$, $C_{57}H_{110}O_6$), food-grade grapeseed oil, (Napa Valley Naturals, Grapeseed Oil, Lot No. S4913 1212, $12.29/750 mL, Supplier: Whole Foods Market, MW: 280.445 g/mol, density 0.90 g/cm$^3$, $C_{18}H_{32}O_2$), castor oil (Thermo Scientific, Stock No. AAL04224AK, CAS 8001-79-4, $28.45/250 mL, Supplier: Thermo Fisher Scientific, MW: 933.45 g/mol, density 0.961 g/cm$^3$, $C_{57}H_{104}O_9$).[16-19]

TABLE 2 Nominal $T_2$ map pulse sequence parameters

| Parameter | 0.35 T | 1.5 T | 3 T |
|---|---|---|---|
| TE | n*20 ms, n=1,32 | n*20 ms, n=1,32 | n*20 ms, n=1,32 |
| TR | 15 s | 15 s | 10 s |
| Flip angle | 90° | 90° | 90° |
| Bandwidth | 130 Hz/px | 128 Hz/px | 128 Hz/px |
| Resolution | 3x3x5 mm$^3$ | 3x3x5 mm$^3$ | 3x3x5 mm$^3$ |

**Permittivities:** Complex permittivity measurements were performed on a benchtop using an Agilent 85070E high-temperature coaxial dielectric probe (Santa Clara, CA) connected to an Agilent E4991A impedance analyzer with a rated range of 10 MHz to 3 GHz.[20] Measurements ranged from 14 MHz to 1 GHz in 2 MHz intervals. A four-point calibration (short, open circuit, 50 ohm load, and low-loss capacitor) of the impedance analyzer was performed. A three-point calibration (air, short, and 21.6$^0$C deionized water) of the dielectric probe was performed after the impedance analyzer was calibrated. The average temperature of the liquids during the permittivity measurements was measured using an Omega Model FOH201 or SAII fiber optic temperature sensor.

Measurements of the hydrogels (Miller's LB agar, gelatin, xanthan gum, sodium polyacrylate, PVA, and PVP) and $MnNO_3$ samples were conducted using the Agilent 85070E probe with an Agilent E5061A network analyzer. A three-point calibration (air, short, and 21.6$^0$C deionized water) of the dielectric probe was performed before measurements. Measurements ranged from 14 MHz to 1.5 GHz in 15 MHz intervals. We extrapolated the static dielectrics ($\varepsilon_s$) and electric conductivities ($\sigma_s$) down to 14 MHz since the dielectric probe is rated for a range from 200 MHz to 20 GHz with the network analyzer.

The complex permittivity data were fit to the Cole-Cole relaxation model using Mathematica to derive the static dielectric constants ($\varepsilon_s$) and static electrical conductivities ($\sigma_s$).[21-24] The static electrical conductivity $\sigma_s$ was calculated from the imaginary component of the complex permittivity ($\varepsilon''$) using:

$$\sigma(\omega) = \varepsilon_0 \omega \varepsilon''(\omega) + \sigma_s \quad (4)$$

where $\varepsilon_0$ is the permittivity of free space.

Linear fits of $\varepsilon_s$ and $\sigma_s$ versus concentration [C] were generated as:

$$\varepsilon_s \approx \varepsilon_{s,0} + a[C] \quad (5)$$

$$\sigma_s \approx \sigma_{s,0} + b[C] \quad (6)$$

where *a* and *b* were the respective linear slopes, and $\varepsilon_{s,0}$ and $\sigma_{s,0}$ were the intercepts.

Linear fits of relaxivities were also calculated for the mixtures based on their concentrations [C] and relaxation rates:

$$R_n = \frac{1}{T_n} = R_{n,0} + r_n[C] \quad (7)$$

where n=1 for $T_1$ and n=2 for $T_2$, respectively. $R_n$ is the relaxation rate, $R_{n,0}$ is the intercept, and $r_n$ is the relaxivity. Additional relaxivity terms can be added to Equation 7 to address the combination of additional ingredients.[25]

The DC electrical conductivity $\sigma_{DC}$ for each concentration was measured using an Oakton CON 110 handheld conductivity/TDS/ºC/ºF meter with RS232. A manual single-point calibration was performed using a potassium chloride 84 µS conductivity standard (LaMotte, Lot No. 1286217, CAS 7447-40-7, $8.95/30 mL, Supplier: Extech Instruments). The average temperature of the liquids during the conductivity measurements was 21.5 +/- 0.1 ºC and was measured using the Oakton CON 110.

**3. RESULTS**

Data plots of the relaxation rates, relaxivities, static dielectrics, static electrical conductivities, and DC conductivities are provided in Figures 1-23. Error bars represent uncertainties in nonlinear fits for $T_1$ and $T_2$ relaxation rates ($R_1$ and $R_2$, respectively) vs. concentration, and linear fits for $T_1$ and $T_2$ relaxivities ($r_1$ or $r_2$, respectively) vs. field strength. Error bars represent uncertainties in linear fits of static dielectrics ($\varepsilon_s$), and static and DC electrical conductivities ($\sigma_s$ and $\sigma_{DC}$, respectively) vs. concentration. Paired t-tests were used to compare $\sigma_s$ and $\sigma_{DC}$. Correlations were calculated between $\varepsilon_s$, $\sigma_s$, and $\sigma_{DC}$. Relaxation times ($T_1$, $T_2$), relaxivities ($r_1$, $r_2$), relaxation rate intercepts ($R_{1,0}$, $R_{2,0}$), static dielectric ($\varepsilon_s$), static conductivity ($\sigma_s$), and DC conductivity ($\sigma_0$) and uncertainties are provided in Tables 3-15.

**Hydrogels:**

The $T_1$ and $T_2$ relaxation rates ($R_1$ and $R_2$, respectively) as a function of Miller's LB agar concentration are shown in Figures 1a and 1b The $T_1$ and $T_2$ relaxivities ($r_1$ and $r_2$, respectively) as a function of field strength are shown in Figures 1c and 1d. Miller's LB agar is a mixture of agar (32% w/w), tryptone (27% w/w), NaCl (27% w/w), and yeast (14% w/w) agar. $T_1$, $T_2$ was consistent with earlier studies when corrected for agar concentration.[26]

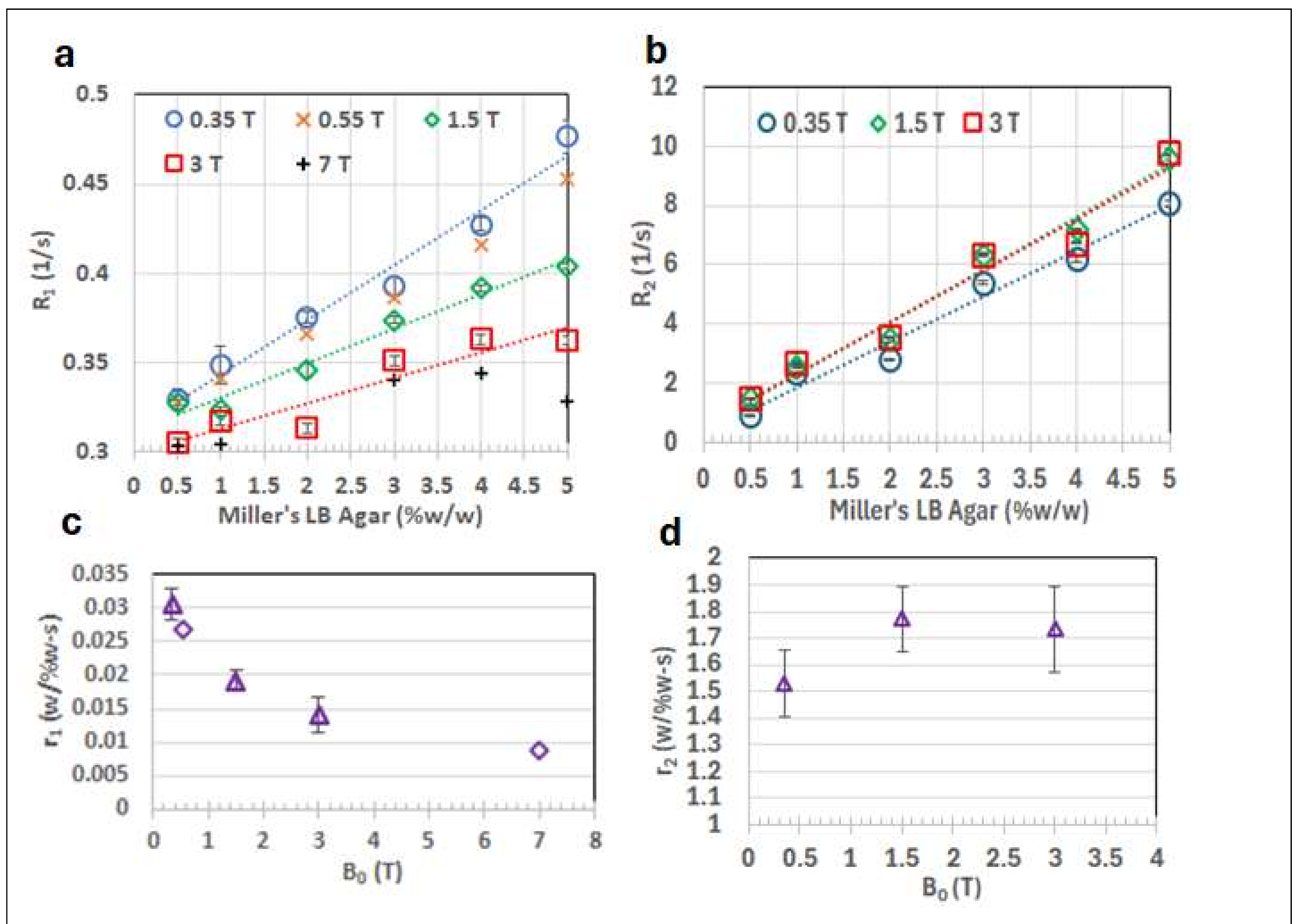

FIGURE 1 Relaxation rates ($R_1$, $R_2$) for different concentrations of Miller's LB agar in water: a) $R_1$ and b) $R_2$, and relaxivities ($r_1$, $r_2$) as a function of field strength for c) $r_1$ and d) $r_2$. Miller's LB agar is a mixture including agar (32% w/w). $R_1$ (×, +) and $r_1$ (◇) interpolations/extrapolations are included for 0.55 T and 7 T. Note the $R_1$ extrapolated for 7 T and concentration 5% w/w is lower than $R_1$ for 4% w/w suggesting that the extrapolation is false.

The dielectric strengths and electrical conductivities for different concentrations of Miller's LB agar are shown in Figure 2. The relaxation times, relaxivities, and permittivity measurements for Miller's LB agar are summarized in Table 3.

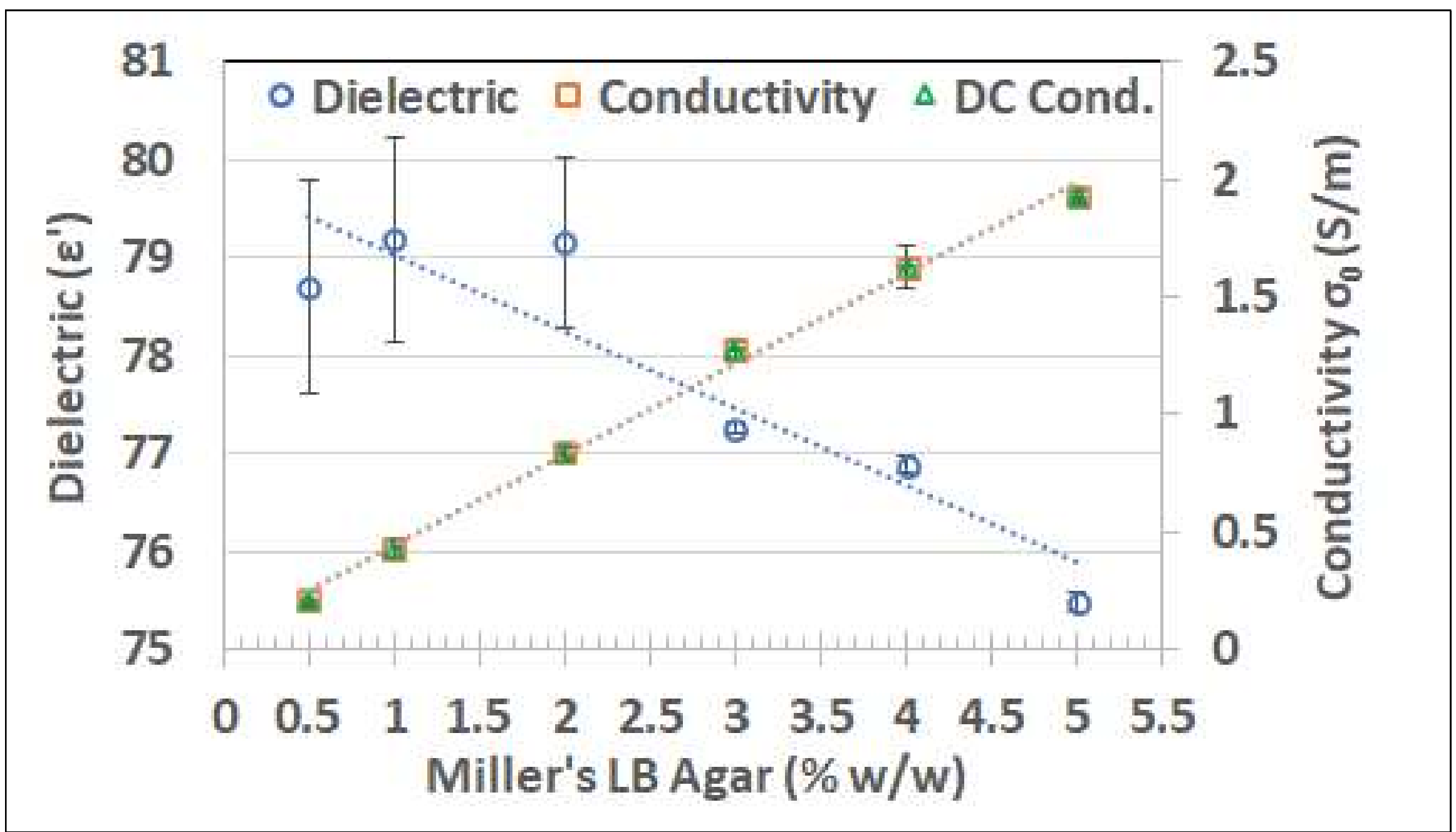

FIGURE 2 Static dielectric constant $\varepsilon_s$ (O, left axis), static electrical conductivity $\sigma_s$ (□) and DC conductivity $\sigma_{DC}$ from the conductivity meter (Δ) are shown for different concentrations of Miller's LB agar in water. $\sigma_s$ and $\sigma_{DC}$ were highly correlated (r=0.997). $\varepsilon_s$ and $\sigma_s$ were highly correlated (r=-0.912).

TABLE 3 Miller's LB agar mixture relaxation times and fit uncertainties in ms measured at room temperature 20°C. Units: $T_1$ and $T_2$ (ms), $r_1$ and $r_2$ (w/%w-s), $R_{1,0}$ and $R_{2,0}$ (1/s), *a* (w/%w), *b* (S/m·w/%w), $\sigma_{s,0}$ (S/m).

| **Concentration (%w/w)** | | **0.35 T (14.71 MHz)** | **1.5 T (63.89 MHz)** | **3 T (123.21 MHz)** |
|---|---|---|---|---|
| 0.5 | $T_1$ | 3036.76 ± 53.45 | 3055.48 ± 10.69 | 3271.60 ± 22.62 |
| | $T_2$ | 1112.92 ± 35.64 | 1155.31 ± 39.89 | 1207.02 ± 14.29 |
| 1 | $T_1$ | 2869.06 ± 84.47 | 3086.54 ± 14.47 | 3150.17 ± 27.86 |
| | $T_2$ | 428.51. ± 5.65 | 396.03 ± 5.14 | 376.29 ± 6.07 |
| 2 | $T_1$ | 2662.90 ± 23.25 | 2889.71 ± 36.00 | 3193.20 ± 28.56 |
| | $T_2$ | 361.56 ± 4.92 | 290.66 ± 3.69 | 283.51 ± 4.23 |
| 3 | $T_1$ | 2546.08 ± 31.48 | 2674.53 ± 11.16 | 2849.65 ± 23.05 |
| | $T_2$ | 185.56 ± 1.20 | 157.88 ± 1.26 | 157.92 ± 2.08 |
| 4 | $T_1$ | 2340.55 ± 19.45 | 2553.39 ± 12.89 | 2752.67 ± 21.40 |
| | $T_2$ | 162.13 ± 12.46 | 139.17 ± 0.93 | 149.16 ± 2.08 |
| 5 | $T_1$ | 2098.00 ± 41.01 | 2475.63 ± 8.83 | 2752.67 ± 18.81 |
| | $T_2$ | 123.92 ± 6.26 | 104.10 ± 0.57 | 102.63 ± 1.80 |
| Relaxivities | $r_1$ | 0.030± 0.00 | 0.02 ± 0.00 | 0.01 ± 0.00 |
| | $R_{1,0}$ | 0.31 ± 0.01 | 0.31 ± 0.00 | 0.30 ± 0.01 |
| | $r_2$ | 1.53 ± 0.13 | 1.77 ± 0.12 | 1.73 ± 0.16 |
| | $R_{2,0}$ | 0.31 ± 0.38 | 0.51 ± 0.37 | 0.59 ± 0.49 |

| Permittivities | | | | |
|---|---|---|---|---|
| Dielectric | | Conductivity | | DC Conductivity |
| $\varepsilon_{s,0}$ | 7.98E1 ± 4.93E-1 | $\sigma_{s,0}$ | 5.10E-2 ± 3.93E-2 | 8.22E-2 ± 3.48E-2 |
| *a* | -7.88E-1 ± 1.62E-1 | *b* | 3.87E-1 ± 1.30E-2 | 3.81E-1 ± 1.15E-2 |

The $T_1$ and $T_2$ relaxation rates ($R_1$ and $R_2$, respectively) as a function of gelatin concentration are shown in Figures 3a and 3b The $T_1$ and $T_2$ relaxivities ($r_1$ and $r_2$, respectively) as a function of field strength are shown in Figures 3c and 3d.

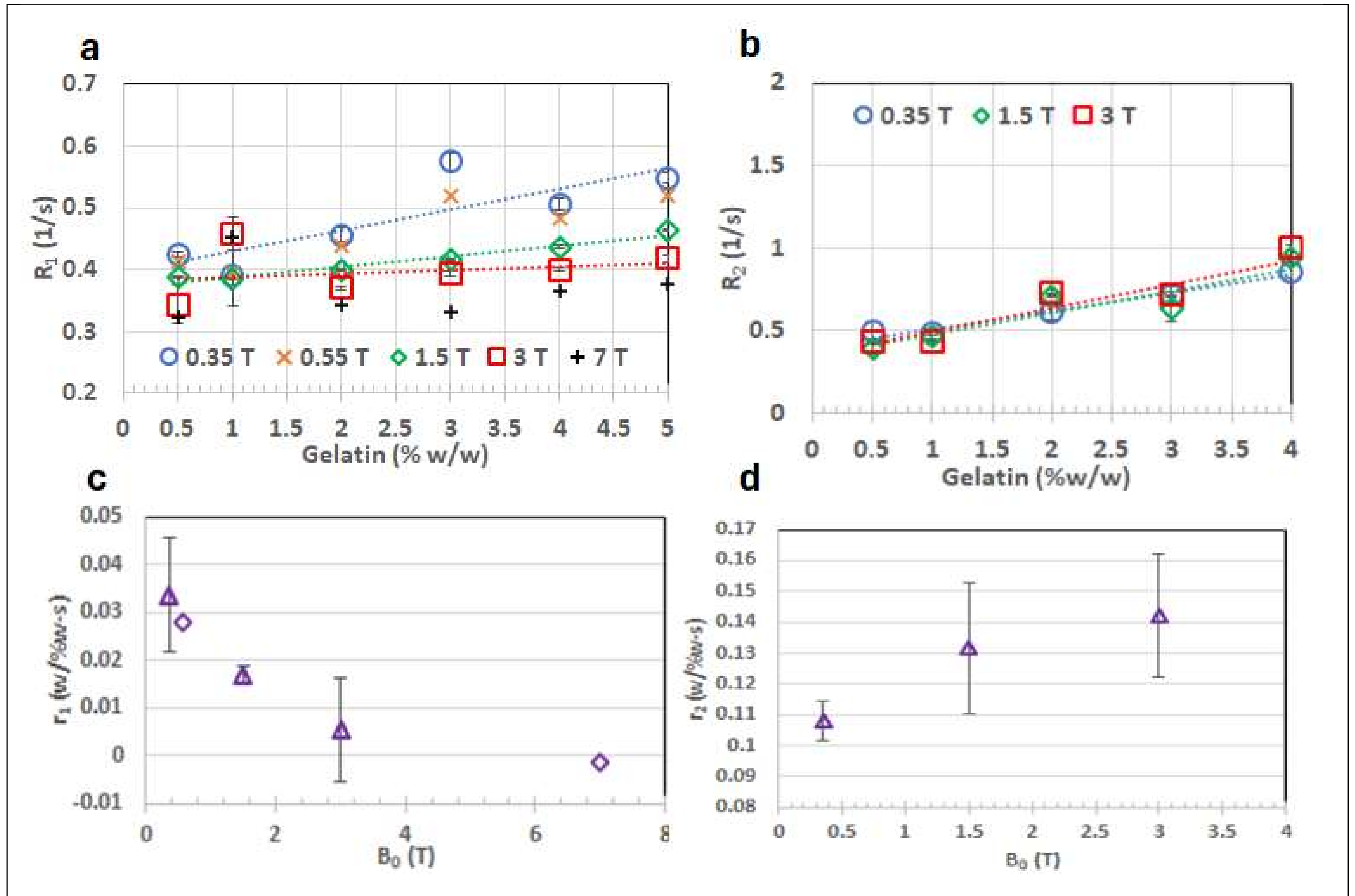

FIGURE 3 Relaxation rates for different concentrations of gelatin: a) $R_1$ and b) $R_2$, and gelatin relaxivities vs. field strength for c) $r_1$ and d) $r_2$. $R_1$ (×, +) and $r_1$ (◇) interpolations/extrapolations are included for 0.55 T and 7 T.

The dielectric strengths and electrical conductivities for different concentrations of gelatin are shown in Figure 4. The relaxation times, relaxivities, and permittivity measurements for gelatin are summarized in Table 4.

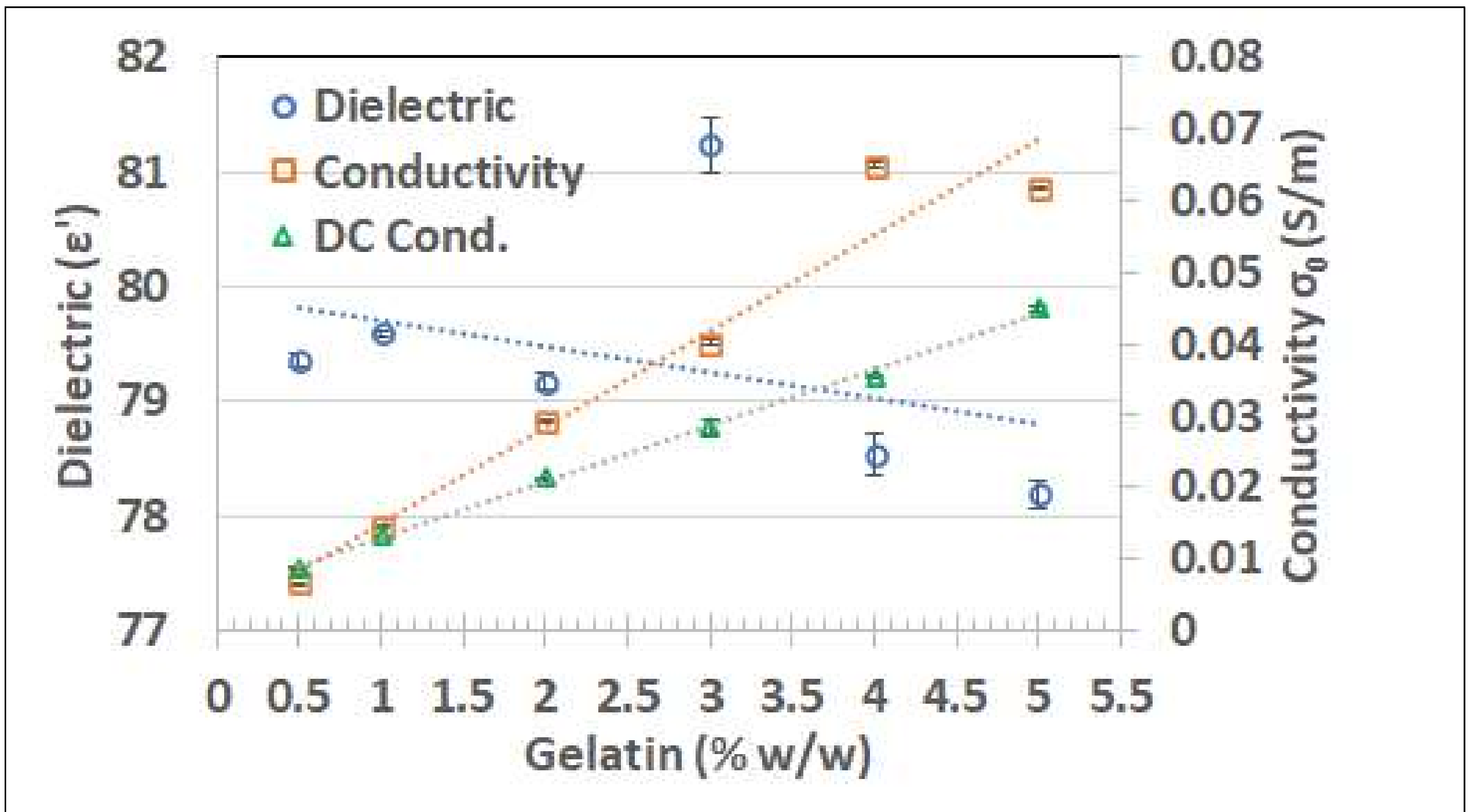

FIGURE 4 Static dielectric constant $\varepsilon_s$ (O, left axis), static electrical conductivity $\sigma_s$ (□) and DC conductivity $\sigma_{DC}$ from the conductivity meter (Δ) are shown for different concentrations of gelatin in water. $\sigma_s$ and $\sigma_{DC}$ were highly correlated (r=0.966).

TABLE 4 Gelatin relaxation times and fit uncertainties in ms measured at room temperature 20°C. Units: $T_1$ and $T_2$ (ms), $r_1$ and $r_2$ (w/%w-s), $R_{1,0}$ and $R_{2,0}$ (1/s), *a* (w/%w), *b* (S/m·w/%w), $\sigma_{s,0}$ (S/m).

| Concentration (%w/w) | | 0.35 T (14.71 MHz) | 1.5 T (63.89 MHz) | 3 T (123.21 MHz) |
|---|---|---|---|---|
| 0.5 | $T_1$ | 2353.39 ± 46.44 | 2578.61 ± 9.26 | 2910.22 ± 24.83 |
| | $T_2$ | 2013.33 ± 254.63 | 2562.23 ± 177.67 | 2203.17 ± 68.38 |
| 1 | $T_1$ | 2553.88 ± 322.93 | 2594.99 ± 10.73 | 2180.51 ± 128.70 |
| | $T_2$ | 2026.55 ± 95.43 | 2134.89 ± 122.63 | 2213.18 ± 71.89 |
| 2 | $T_1$ | 2191.25 ± 54.01 | 2502.77± 8.69 | 2701.10 ± 20.67 |
| | $T_2$ | 1639.22 ± 123.04 | 1400.78 ± 18.10 | 1343.08 ± 29.57 |
| 3 | $T_1$ | 1739.71 ± 40.90 | 2401.19 ± 8.41 | 2542.23 ± 27.17 |
| | $T_2$ | 1395.19 ± 92.56 | 1569.98 ± 193.79 | 1363.78 ± 31.45 |
| 4 | $T_1$ | 1977.73 ± 35.51 | 2292.92 ± 12.54 | 2499.97 ± 15.20 |
| | $T_2$ | 1166.15 ± 45.58 | 1063.30 ± 23.56 | 988.80 ± 12.49 |
| 5 | $T_1$ | 1821.11 ± 24.43 | 2159.51 ± 8.98 | 2395.55 ± 26.14 |
| | $T_2$ | 1049.05 ± 22.85 | 1017.94 ± 6.19 | 956.20 ± 16.03 |
| Relaxivities | $r_1$ | 0.03 ± 0.01 | 0.02 ± 0.00 | 0.01 ± 0.01 |
| | $R_{1,0}$ | 0.40 ± 0.04 | 0.37 ± 0.01 | 0.38 ± 0.03 |
| | $r_2$ | 0.11 ± 0.01 | 0.13 ± 0.02 | 0.14 ± 0.20 |

| | $R_{2,0}$ | 0.41 ± 0.02 | 0.35 ± 0.06 | 0.36 ± 0.06 |
|---|---|---|---|---|
| Permittivities | | | | |
| Static Dielectric | | Static Conductivity | | DC Conductivity |
| $\varepsilon_{s,0}$ | 7.99E1 ± 8.57E-1 | $\sigma_{s,0}$ | 1.68E-3 ± 4.71E-3 | 5.11E-3 ± 5.11E-4 |
| *a* | -2.28E-1 ± 2.82E-1 | *b* | 1.34E-2 ± 1.55E-3 | 7.82E-3 ± 1.69E-4 |

The $T_1$ and $T_2$ relaxation rates ($R_1$ and $R_2$, respectively) as a function of polyethylene glycol (PEG) concentration are shown in Figures 5a and 5b The $T_1$ and $T_2$ relaxivities ($r_1$ and $r_2$, respectively) as a function of field strength are shown in Figures 5c and 5d.

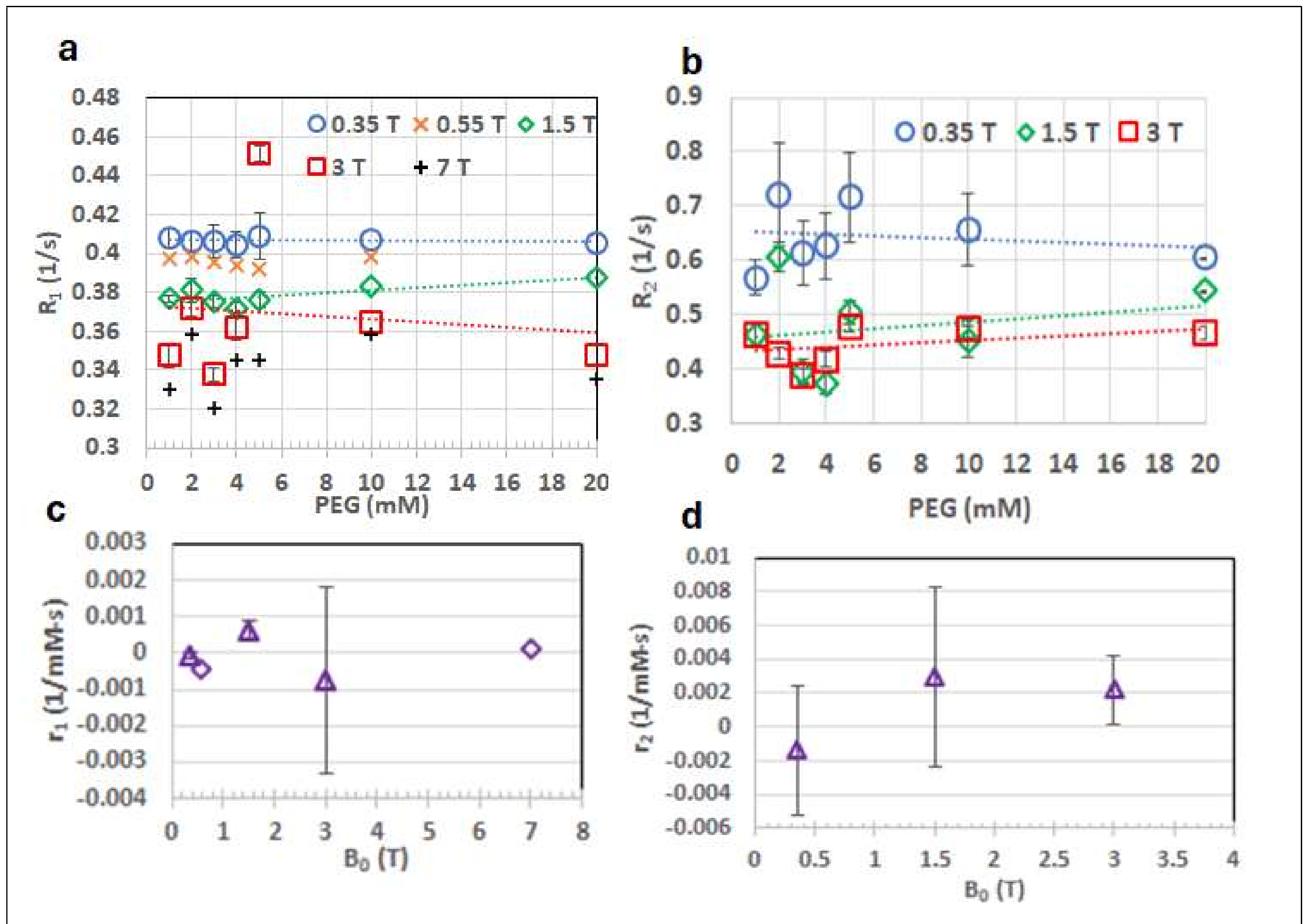

FIGURE 5 Relaxation rates for different concentrations of polyethylene glycol (PEG): a) $R_1$ and b) $R_2$, and PEG relaxivities vs. field strength for c) $r_1$ and d) $r_2$. $R_1$ (×, +) and $r_1$ (◇) interpolations/extrapolations are included for 0.55 T and 7 T.

The dielectric strengths and electrical conductivities for different concentrations of PEG are shown in Figure 6. The relaxation times, relaxivities, and permittivity measurements for PEG are summarized in Table 5.

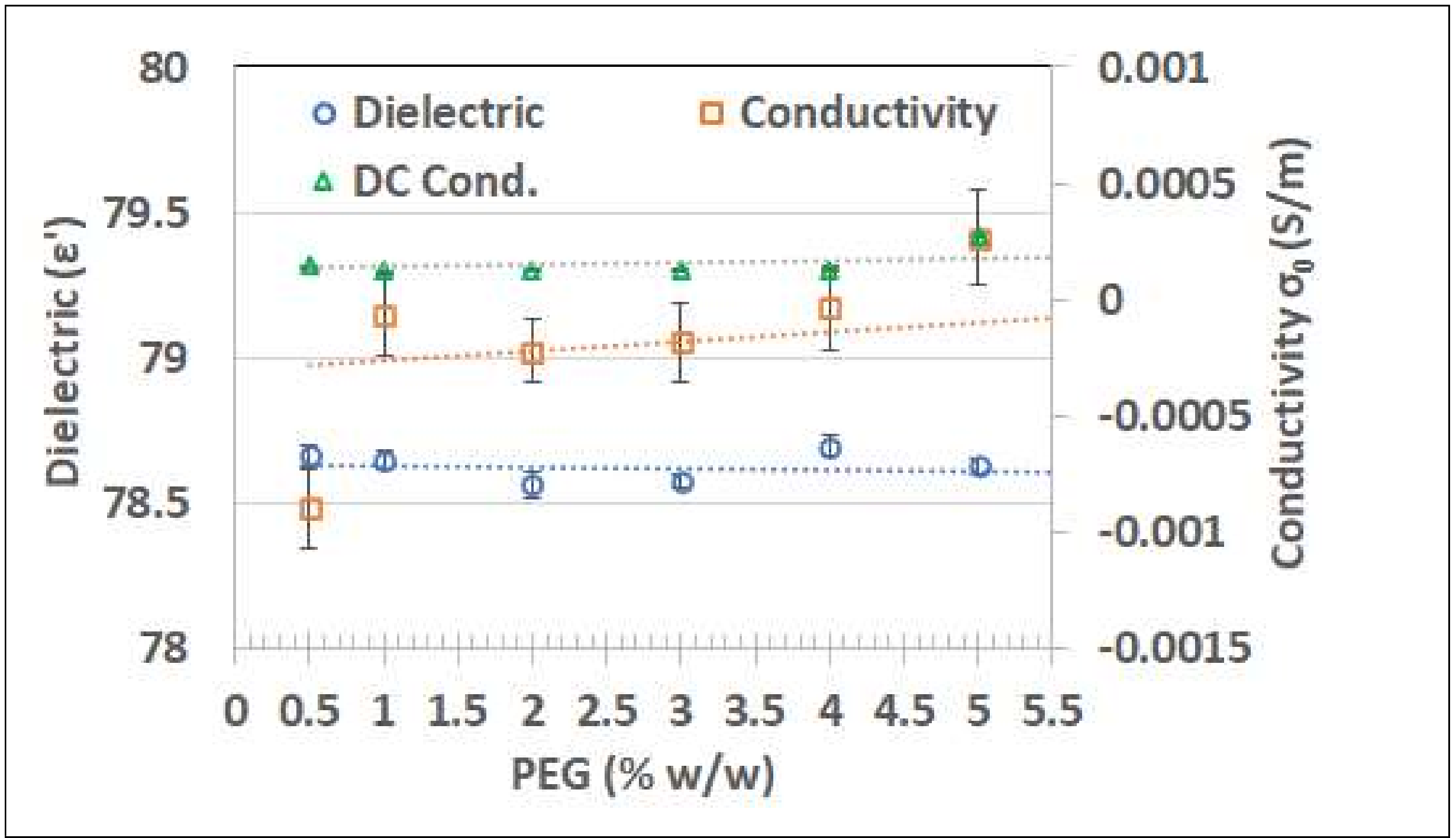

FIGURE 6 Static dielectric constant $\varepsilon_s$ (○, left axis), static electrical conductivity $\sigma_s$ (□) and DC conductivity $\sigma_{DC}$ from the conductivity meter (Δ) are shown for different concentrations of PEG in water. $\sigma_s$ and $\sigma_{DC}$ were correlated (r=0.682).

TABLE 5 PEG relaxation times and fit uncertainties in ms measured at room temperature 20°C. Units: $T_1$ and $T_2$ (ms), $r_1$ and $r_2$ (1/mM-s), $R_{1,0}$ and $R_{2,0}$ (1/s), *a* (1/mM), *b* (S/m·mM), $\sigma_{s,0}$ (S/m).

| **Concentration (mM)** | | **0.35 T (14.71 MHz)** | **1.5 T (63.89 MHz)** | **3 T (123.21 MHz)** |
|---|---|---|---|---|
| 1 | $T_1$ | 2451.57 ± 15.39 | 2654.44 ± 10.64 | 2877.36 ± 38.25 |
| | $T_2$ | 1760.73 ± 97.93 | 2152.27 ± 98.11 | 2175.70 ± 85.22 |
| 2 | $T_1$ | 2461.77 ± 18.78 | 2619.44 ± 7.55 | 2692.26 ± 100.80 |
| | $T_2$ | 1385.41 ± 183.30 | 1646.89 ± 73.65 | 2297.23 ± 46.03 |
| 3 | $T_1$ | 2462.23 ± 20.87 | 2662.66 ± 8.20 | 2958.85 ± 48.90 |
| | $T_2$ | 1628.87 ± 158.45 | 2522.98 ± 144.89 | 2499.99 ± 75.04 |
| 4 | $T_1$ | 2472.98 ± 20.35 | 2690.55 ± 10.13 | 2759.24 ± 66.67 |
| | $T_2$ | 1595.33 ± 152.45 | 2665.09 ± 147.51 | 2327.21 ± 58.87 |
| 5 | $T_1$ | 2447.27 ± 19.81 | 2657.64 ± 9.46 | 2215.34 ± 111.17 |
| | $T_2$ | 1395.39 ± 160.76 | 1981.40 ± 79.54 | 2085.52 ± 31.06 |
| 10 | $T_1$ | 2457.10 ± 17.60 | 2610.58 ± 10.53 | 2741.32 ± 33.84 |
| | $T_2$ | 1520.81 ± 151.64 | 2218.64 ± 139.03 | 2086.51 ± 79.95 |
| 20 | $T_1$ | 2465.77 ± 30.45 | 2579.28 ± 6.79 | 2877.65 ± 14.25 |
| | $T_2$ | 1652.47 ± 79.40 | 1836.97 ± 75.37 | 2102.10 ± 44.60 |
| Relaxivities | $r_1$ | 0.00 ± 0.00 | 0.00 ± 0.00 | 0.00 ± 0.00 |
| | $R_{1,0}$ | 0.41 ± 0.00 | 0.37 ± 0.00 | 0.37 ± 0.02 |

| | $r_2$ | 0.00 ± 0.00 | 0.00 ± 0.01 | 0.00 ± 0.00 |
|---|---|---|---|---|
| | $R_{2,0}$ | 0.65 ± 0.03 | 0.46 ± 0.05 | 0.43 ± 0.02 |
| Permittivities | | | | |
| Static Dielectric | | Static Conductivity | | DC Conductivity |
| $\varepsilon_{s,0}$ | 7.86E1 ± 3.03E-2 | $\sigma_{s,0}$ | -2.99E-4 ± 1.55E-4 | 1.36E-4 ± 2.58E-5 |
| *a* | -4.69E-3 ± 3.63E-3 | *b* | 4.07E-5 ± 1.86E-5 | 8.72E-6 ± 3.10E-6 |

The $T_1$ and $T_2$ relaxation rates ($R_1$ and $R_2$, respectively) as a function of polyvinyl alcohol (PVA) concentration are shown in Figures 7a and 7b The $T_1$ and $T_2$ relaxivities ($r_1$ and $r_2$, respectively) as a function of field strength are shown in Figures 7c and 7d.

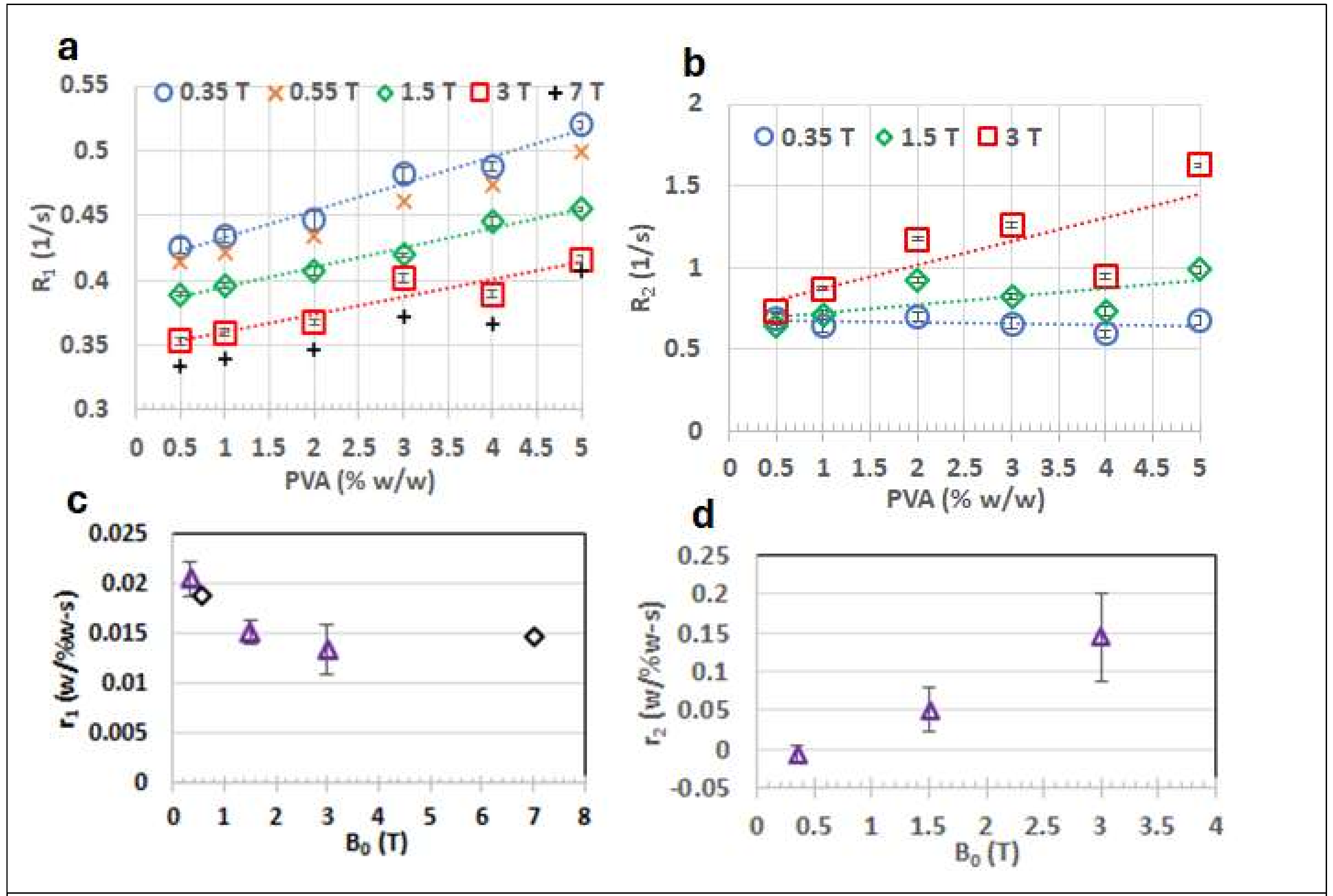

FIGURE 7 Relaxation rates for different concentrations of polyvinyl alcohol (PVA): a) $R_1$ and b) $R_2$, and PVA relaxivities vs. field strength for c) $r_1$ and d) $r_2$. $R_1$ (×, +) and $r_1$ (◇) interpolations/extrapolations are included for 0.55 T and 7 T.

The dielectric strengths and electrical conductivities for different concentrations of PVA are shown in Figure 8. The relaxation times, relaxivities, and permittivity measurements for PVA are summarized in Table 6.

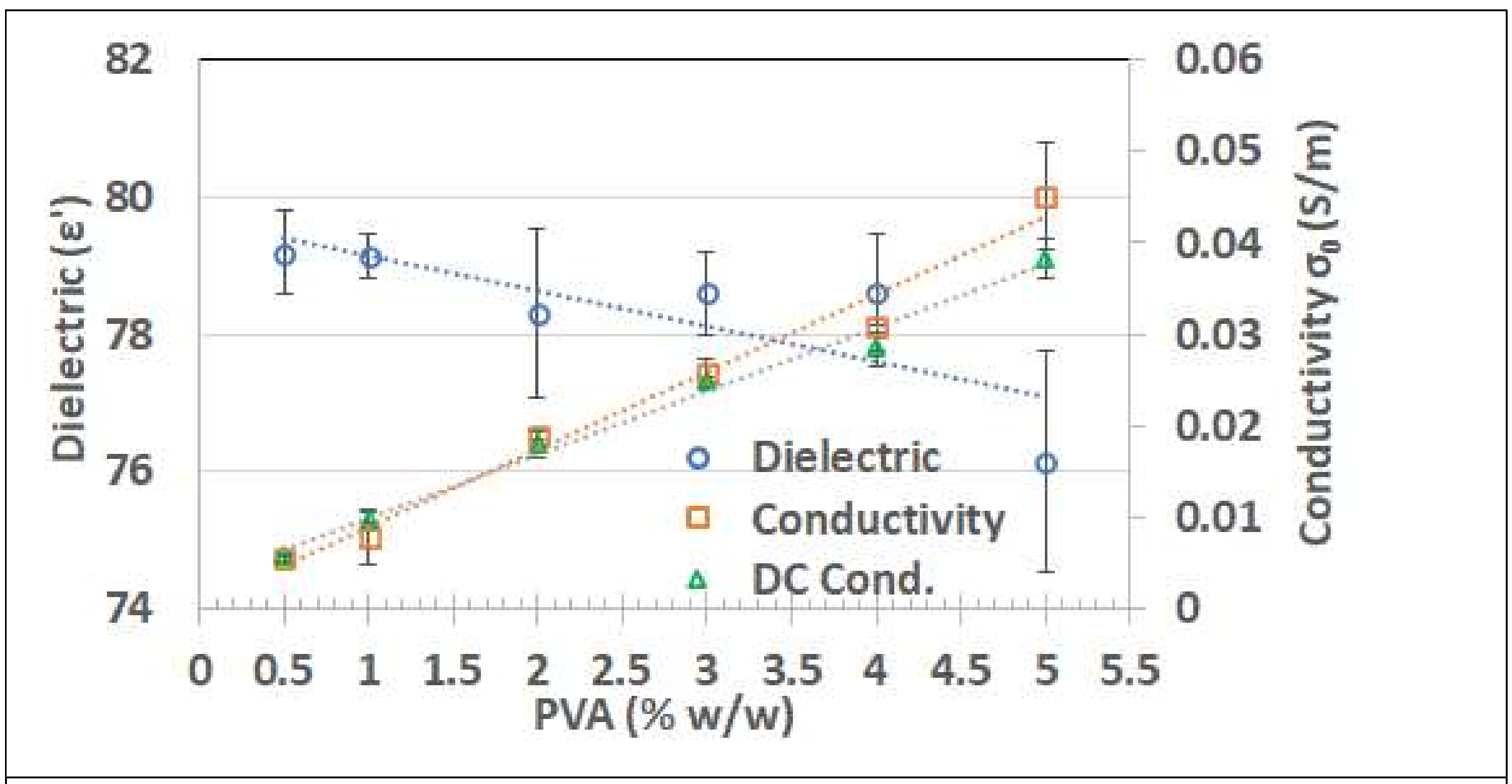

FIGURE 8 Static dielectric constant $\varepsilon_s$ (○, left axis), static electrical conductivity $\sigma_s$ (□) and DC conductivity $\sigma_{DC}$ from the conductivity meter (Δ) are shown for different concentrations of PVA in water. $\sigma_s$ and $\sigma_{DC}$ were highly correlated (r=0.996). $\varepsilon_s$ and $\sigma_s$ were highly correlated (r=-0.870).

TABLE 6 PVA relaxation times and fit uncertainties in ms measured at room temperature 20°C. Units: $T_1$ and $T_2$ (ms), $r_1$ and $r_2$ (w/%w-s), $R_{1,0}$ and $R_{2,0}$ (1/s), *a* (w/%w), *b* (S/m·w/%w), $\sigma_{s,0}$ (S/m).

| **Concentration (%w/w)** | | **0.35 T (14.71 MHz)** | **1.5 T (63.89 MHz)** | **3 T (123.21 MHz)** |
|---|---|---|---|---|
| 0.5 | $T_1$ | 2346.60 ± 28.88 | 2568.33 ± 10.74 | 2829.28 ± 23.50 |
| | $T_2$ | 1462.93 ± 80.57 | 1546.75 ± 60.35 | 1355.41 ± 31.40 |
| 1 | $T_1$ | 2301.15 ± 25.93 | 2526.82 ± 11.59 | 2781.07 ± 24.18 |
| | $T_2$ | 1551.56 ± 100.09 | 1410.71 ± 53.60 | 1147.34 ± 9.17 |
| 2 | $T_1$ | 2240.15 ± 37.56 | 2457.38 ± 21.00 | 2716.44 ± 17.02 |
| | $T_2$ | 1433.32 ± 57.80 | 1082.51 ± 15.88 | 853.67 ± 5.73 |
| 3 | $T_1$ | 2076.42 ± 24.92 | 2384.55 ± 10.67 | 2488.19 ± 22.26 |
| | $T_2$ | 1522.31 ± 81.15 | 1214.93 ± 28.82 | 796.51 ± 7.77 |
| 4 | $T_1$ | 2051.60 ± 15.38 | 2242.25 ± 18.63 | 2569.04 ± 20.09 |
| | $T_2$ | 1680.19 ± 67.68 | 1359.11 ± 49.23 | 1058.64 ± 12.60 |
| 5 | $T_1$ | 1922.64 ± 9.75 | 2198.22 ± 7.29 | 2402.47 ± 18.77 |
| | $T_2$ | 1477.18 ± 61.05 | 1010.65 ± 23.13 | 625.23 ± 4.91 |
| Relaxivities | $r_1$ | 0.02 ± 0.00 | 0.02 ± 0.00 | 0.01 ± 0.00 |
| | $R_{1,0}$ | 0.41 ± 0.01 | 0.38 ± 0.00 | 0.35 ± 0.01 |
| | $r_2$ | -0.01 ± 0.01 | 0.05 ± 0.03 | 0.14 ± 0.06 |
| | $R_{2,0}$ | 0.68 ± 0.03 | 0.67 ± 0.09 | 0.73 ± 0.17 |
| Permittivities | | | | |

| Static Dielectric | | Static Conductivity | | DC Conductivity |
|---|---|---|---|---|
| $\varepsilon_{s,0}$ | 7.97E1 ± 5.87E-1 | $\sigma_{s,0}$ | 5.82E-4 ± 1.81E-3 | 3.10E-3 ± 1.07E-3 |
| *a* | -5.17E-1 ± 1.93E-1 | *b* | 8.43E-3 ± 5.95E-4 | 6.91E-3 ± 3.51E-4 |

The $T_1$ and $T_2$ relaxation rates ($R_1$ and $R_2$, respectively) as a function of polyvinylpyrrolidone (PVP) concentration are shown in Figures 9a and 9b The $T_1$ and $T_2$ relaxivities ($r_1$ and $r_2$, respectively) as a function of field strength are shown in Figures 9c and 9d.

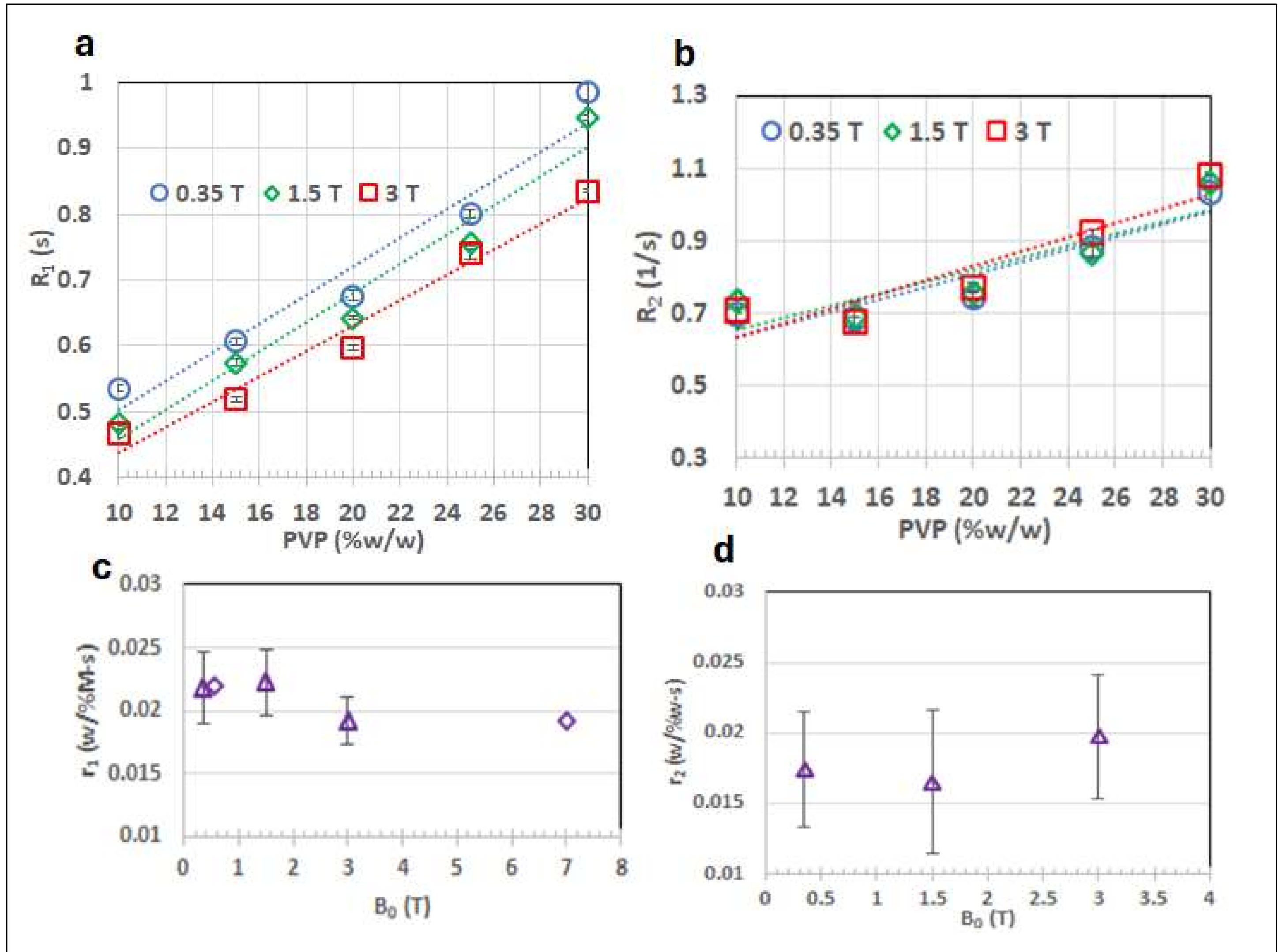

FIGURE 9 Relaxation rates for different concentrations of polyvinylpyrrolidone (PVP): a) $R_1$ and b) $R_2$, and PVP relaxivities vs. field strength for c) $r_1$ and d) $r_2$. $R_1$ (×, +) and $r_1$ (◇) interpolations/extrapolations are included for 0.55 T and 7 T.

The dielectric strengths and electrical conductivities for different concentrations of PVP are shown in Figure 10.

The relaxation times, relaxivities, and permittivity measurements for PVP are summarized in Table 7.

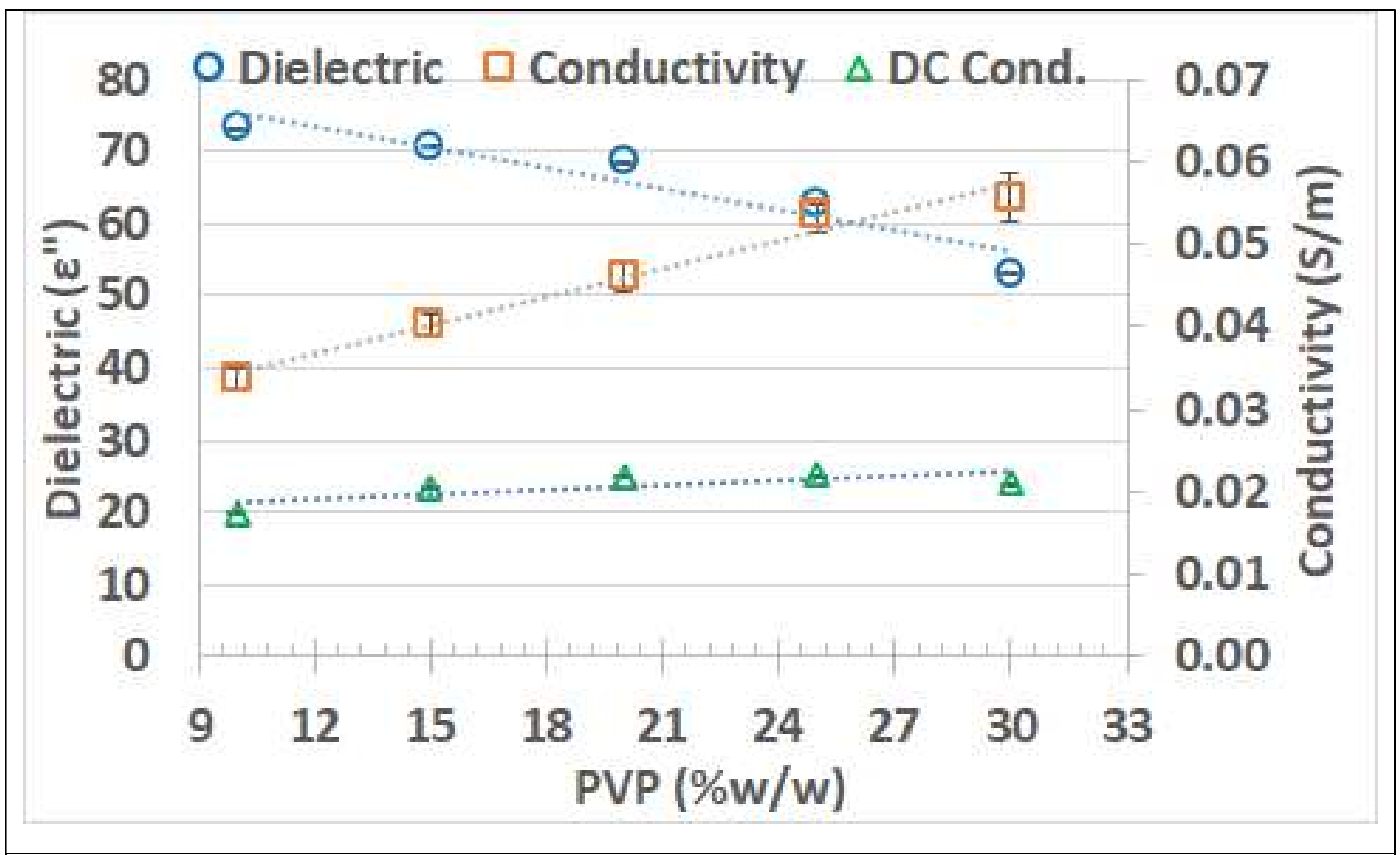

FIGURE 10 Static dielectric constant $\varepsilon_s$ (○, left axis), static electrical conductivity $\sigma_s$ (□) and DC conductivity $\sigma_{DC}$ from the conductivity meter (Δ) are shown for different concentrations of PVP in water. $\sigma_s$ and $\sigma_{DC}$ were highly correlated (r=-0.811) but significantly different ($p<0.01$). $\varepsilon_s$ and $\sigma_s$ were highly correlated (r=-0.811).

TABLE 7 PVP Relaxation times and fit uncertainties in ms measured at room temperature 20°C. Units: $T_1$ and $T_2$ (ms), $r_1$ and $r_2$ (w/%w-s), $R_{1,0}$ and $R_{2,0}$ (1/s), *a* (w/%w), *b* (S/m·w/%w), $\sigma_{s,0}$ (S/m).

| Concentration (%w/w) | | 0.35 T (14.71 MHz) | 1.5 T (63.89 MHz) | 3 T (123.21 MHz) |
|---|---|---|---|---|
| 10 | $T_1$ | 1867.45 ± 16.04 | 2077.23 ± 8.71 | 2146.40 ± 27.68 |
| | $T_2$ | 1435.95 ± 80.32 | 1359.17 ± 46.13 | 1429.87 ± 13.73 |
| 15 | $T_1$ | 1647.79 ± 13.54 | 1740.09 ± 15.93 | 1932.05 ± 13.56 |
| | $T_2$ | 1461.10 ± 44.90 | 1450.21 ± 82.07 | 1481.08 ± 28.67 |
| 20 | $T_1$ | 1478.96 ± 18.45 | 1556.21 ± 7.39 | 1677.32 ± 10.41 |
| | $T_2$ | 1347.76 ± 37.36 | 1321.26 ± 30.61 | 1294.94 ± 21.24 |
| 25 | $T_1$ | 1249.93 ± 9.85 | 1323.69 ± 5.22 | 1352.30 ± 12.51 |
| | $T_2$ | 1132.60± 32.00 | 1153.72 ± 19.52 | 1087.10 ± 13.45 |
| 30 | $T_1$ | 1014.23 ± 12.32 | 1055.45 ± 4.06 | 1196.52 ± 5.39 |
| | $T_2$ | 968.13 ± 16.15 | 943.74 ± 14.72 | 928.17 ± 9.68 |
| Relaxivities | $r_1$ | 0.02 ± 0.00 | 0.02 ± 0.00 | 0.02 ± 0.00 |
| | $R_{1,0}$ | 0.28 ± 0.06 | 0.24 ± 0.06 | 0.25 ± 0.04 |
| | $r_2$ | 0.02 ± 0.00 | 0.02 ± 0.01 | 0.02 ± 0.00 |
| | $R_{2,0}$ | 0.46 ± 0.09 | 0.49 ± 0.11 | 0.44 ± 0.09 |
| Permittivities | | | | |

| Static Dielectric | | Static Conductivity | | DC Conductivity |
|---|---|---|---|---|
| $\varepsilon_{s,0}$ | 8.49E1 ± 3.91E0 | $\sigma_{s,0}$ | 2.29E-2 ± 2.00E-3 | 1.71E-2 ± 1.90E-3 |
| *a* | -9.62E-1 ± 1.84E-1 | *b* | 1.15E-3 ± 9.41E-5 | 1.76E-4 ± 8.94E-5 |

The $T_1$ and $T_2$ relaxation rates ($R_1$ and $R_2$, respectively) as a function of sodium alginate concentration are shown in Figures 11a and 11b The $T_1$ and $T_2$ relaxivities ($r_1$ and $r_2$, respectively) as a function of field strength are shown in Figures 11c and 11d.

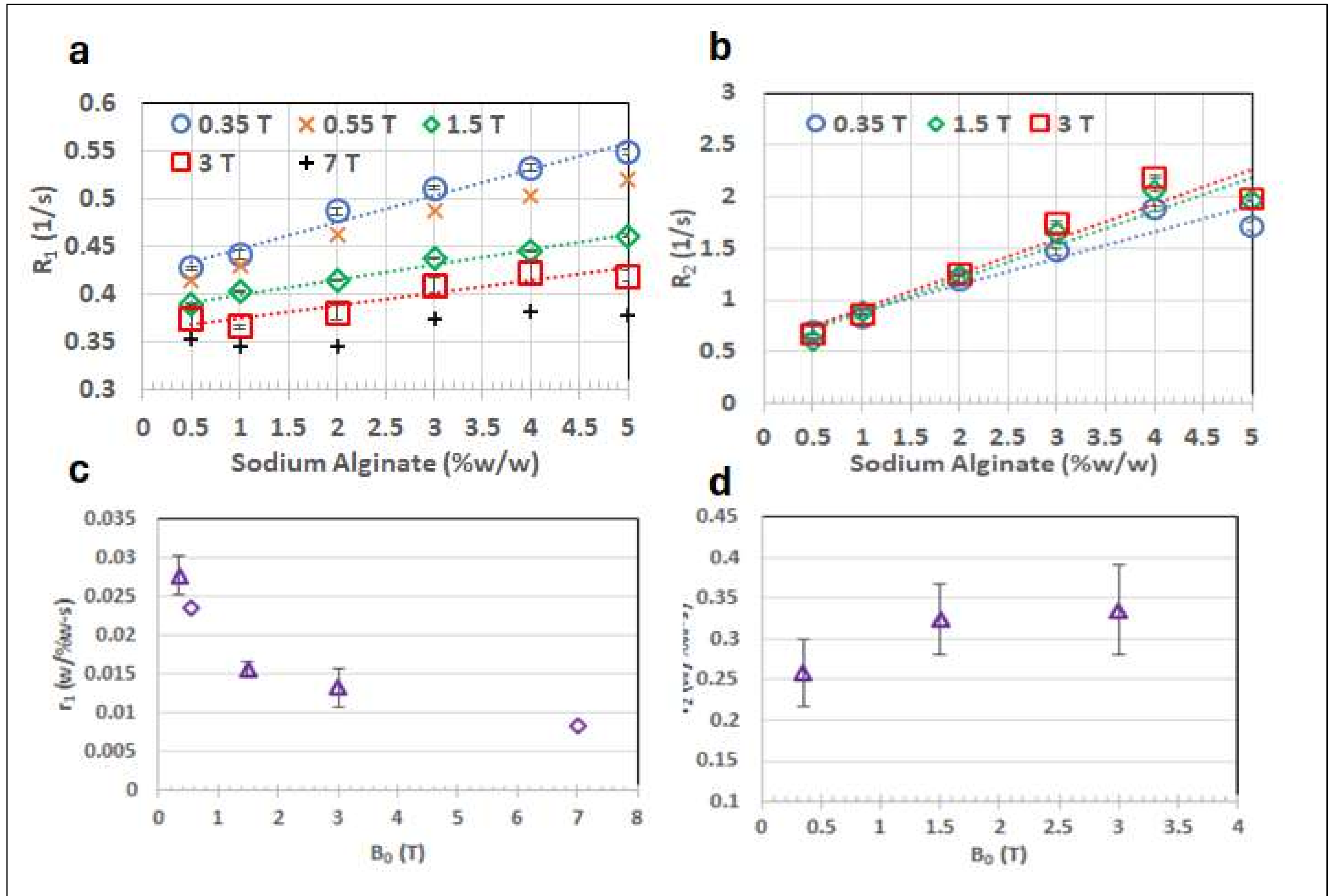

FIGURE 11 Relaxation rates for different concentrations of sodium alginate: a) $R_1$ and b) $R_2$, and sodium alginate relaxivities vs. field strength for c) $r_1$ and d) $r_2$. $R_1$ (×, +) and $r_1$ (◇) interpolations/extrapolations are included for 0.55 T and 7 T.

The dielectric strengths and electrical conductivities for different concentrations of sodium alginate are shown in Figure 12. The relaxation times, relaxivities, and permittivity measurements for sodium alginate are summarized in Table 8.

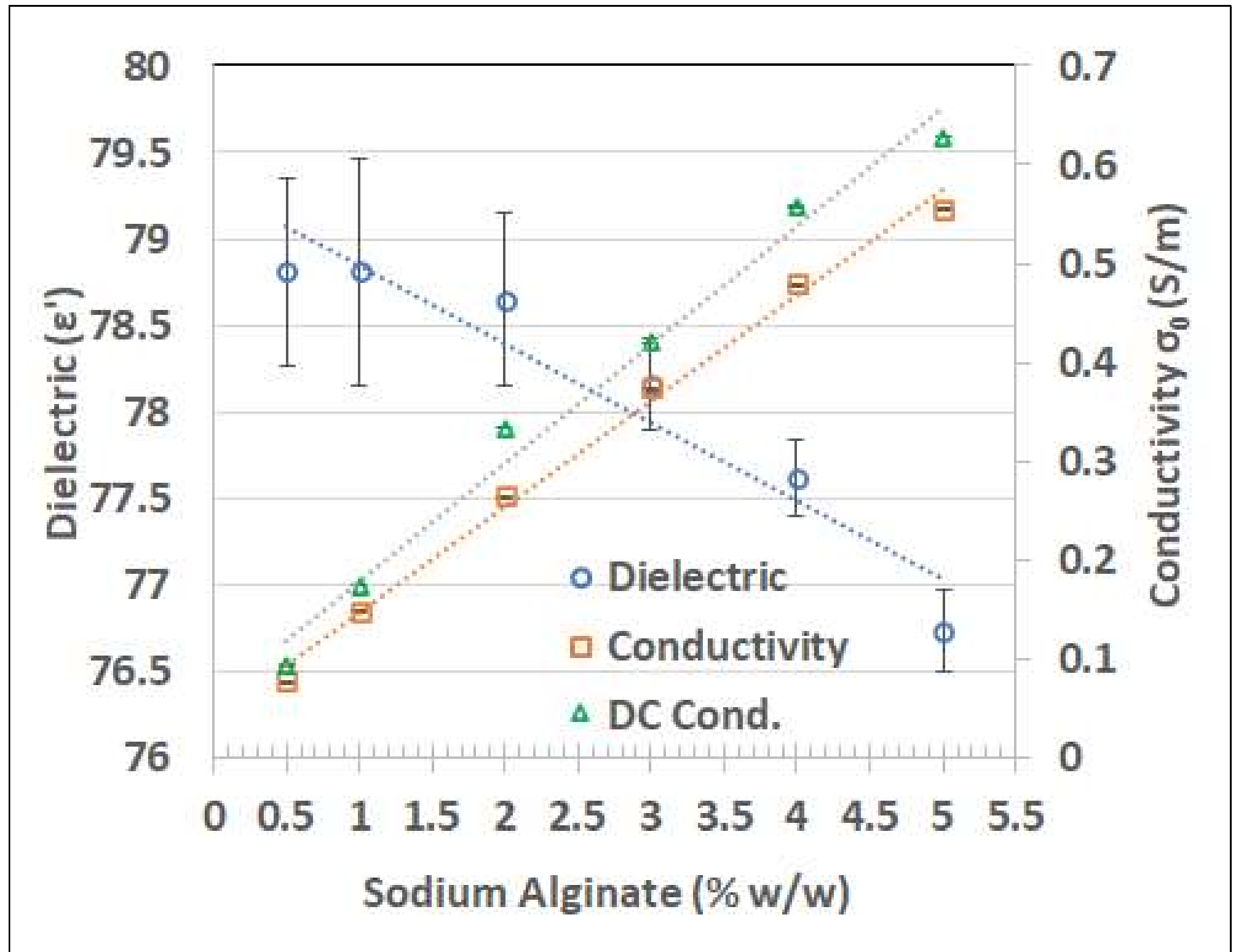

FIGURE 12 Static dielectric constant $\varepsilon_s$ (○, left axis), static electrical conductivity $\sigma_s$ (□) and DC conductivity $\sigma_{DC}$ from the conductivity meter (Δ) are shown for different concentrations of sodium alginate in water. $\sigma_s$ and $\sigma_{DC}$ were highly correlated (r=0.998) and significantly different ($p<0.01$). $\varepsilon_s$ and $\sigma_s$ were highly correlated (r=-0.931).

TABLE 8 Sodium alginate relaxation times and fit uncertainties in ms measured at room temperature 20°C. Units: $T_1$ and $T_2$ (ms), $r_1$ and $r_2$ (w/%w-s), $R_{1,0}$ and $R_{2,0}$ (1/s), *a* (w/%w), *b* (S/m·w/%w), $\sigma_{s,0}$ (S/m).

| **Concentration (%w/w)** | | **0.35 T (14.71 MHz)** | **1.5 T (63.89 MHz)** | **3 T (123.21 MHz)** |
|---|---|---|---|---|
| 0.5 | $T_1$ | 2338.54 ± 12.69 | 2568.82 ± 7.13 | 2678.16 ± 65.82 |
| | $T_2$ | 1415.61 ± 71.43 | 1628.03 ± 53.96 | 1496.10 ± 25.88 |
| 1 | $T_1$ | 2260.97 ± 24.84 | 2485.53 ± 6.99 | 2732.23 ± 13.82 |
| | $T_2$ | 1197.80 ± 22.65 | 1127.44 ± 16.03 | 1144.77 ± 14.94 |
| 2 | $T_1$ | 2051.83 ± 14.36 | 2414.58 ± 5.18 | 2626.46 ± 47.17 |
| | $T_2$ | 840.36 ± 34.24 | 809.01 ± 6.54 | 782.93 ± 6.18 |
| 3 | $T_1$ | 1953.39 ± 7.00 | 2284.79 ± 7.91 | 2442.97 ± 51.60 |
| | $T_2$ | 681.28 ± 15.44 | 609.76 ± 4.02 | 569.29 ± 4.77 |
| 4 | $T_1$ | 1876.86 ± 12.26 | 2245.70 ± 3.14 | 2363.98 ± 71.47 |
| | $T_2$ | 531.04 ± 8.52 | 485.35 ± 3.81 | 460.28 ± 3.76 |
| 5 | $T_1$ | 1818.54 ± 9.38 | 2167.71 ± 7.55 | 2387.52 ± 35.11 |
| | $T_2$ | 585.09 ± 11.80 | 509.82 ± 3.38 | 503.63 ± 2.92 |
| Relaxivities | $r_1$ | 0.03 ± 0.00 | 0.02 ± 0.00 | 0.01 ± 0.00 |
| | $R_{1,0}$ | 0.42 ± 0.01 | 0.39 ± 0.00 | 0.36 ± 0.01 |
| | $r_2$ | 0.26± 0.04 | 0.32 ± 0.04 | 0.34 ± 0.05 |
| | $R_{2,0}$ | 0.63 ± 0.17 | 0.56 ± 0.13 | 0.58 ± 0.17 |

| Permittivities | | | | |
|---|---|---|---|---|
| Static Dielectric | | Static Conductivity | | DC Conductivity |
| $\varepsilon_{s,0}$ | 7.93E1 ± 1.08E0 | $\sigma_{s,0}$ | 4.09E-2 ± 9.25E-3 | 5.84E-2 ± 2.20E-2 |
| *a* | -4.53E-1 ± 3.55E-1 | *b* | 1.07E-1 ± 3.05E-3 | 1.20E-1 ± 7.25E-3 |

The $T_1$ and $T_2$ relaxation rates ($R_1$ and $R_2$, respectively) as a function of sodium polyacrylate concentration are shown in Figures 13a and 13b The $T_1$ and $T_2$ relaxivities ($r_1$ and $r_2$, respectively) as a function of field strength are shown in Figures 13c and 13d.

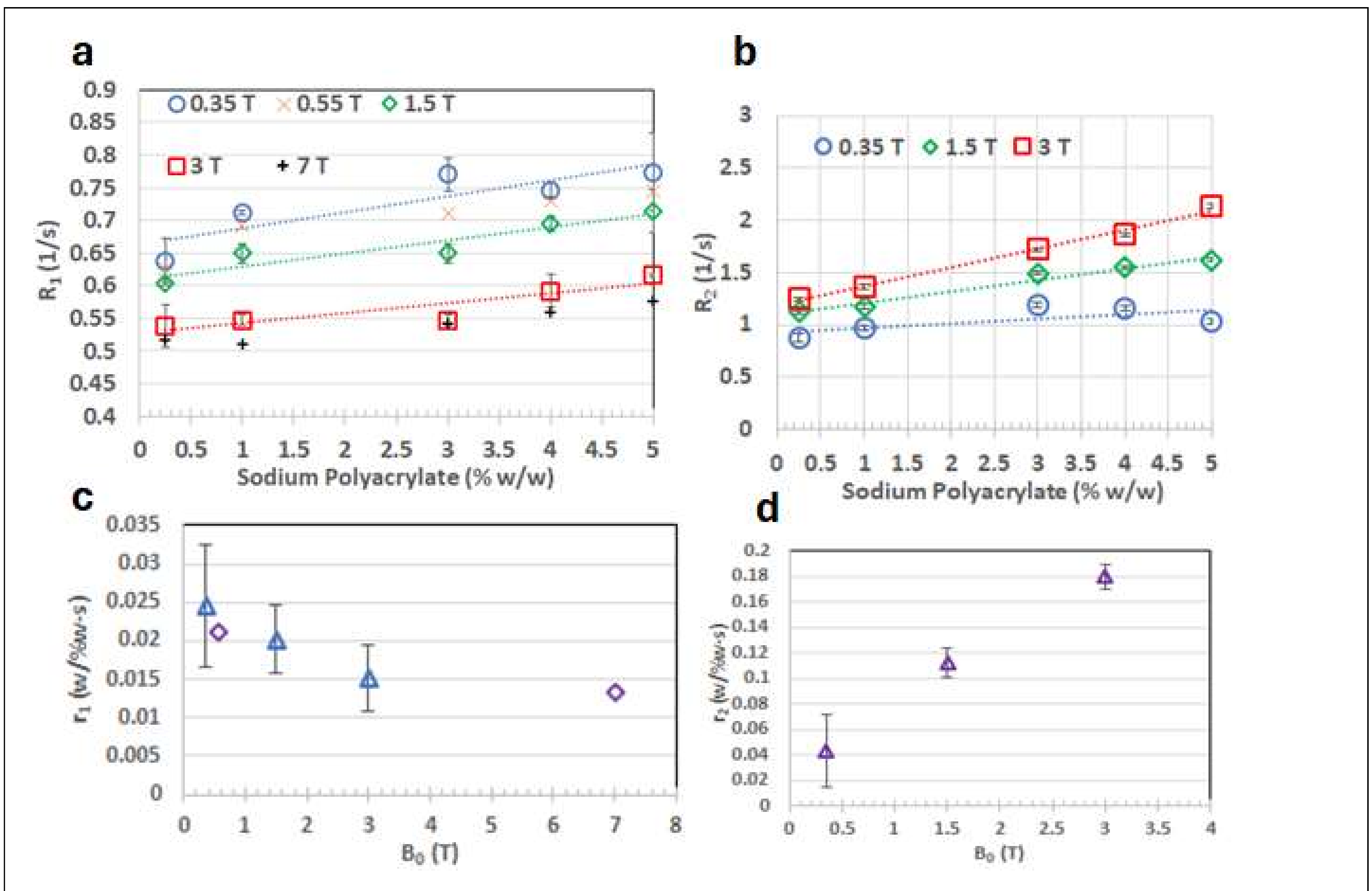

FIGURE 13 Relaxation rates for different concentrations of sodium polyacrylate: a) $R_1$ and b) $R_2$, and sodium polyacrylate relaxivities vs. field strength for c) $r_1$ and d) $r_2$. $R_1$ (×, +) and $r_1$ (◊) interpolations/extrapolations are included for 0.55 T and 7 T.

The dielectric strengths and electrical conductivities for different concentrations of sodium polyacrylate are shown in Figure 14. The relaxation times, relaxivities, and permittivity measurements for sodium polyacrylate are summarized in Table 9.

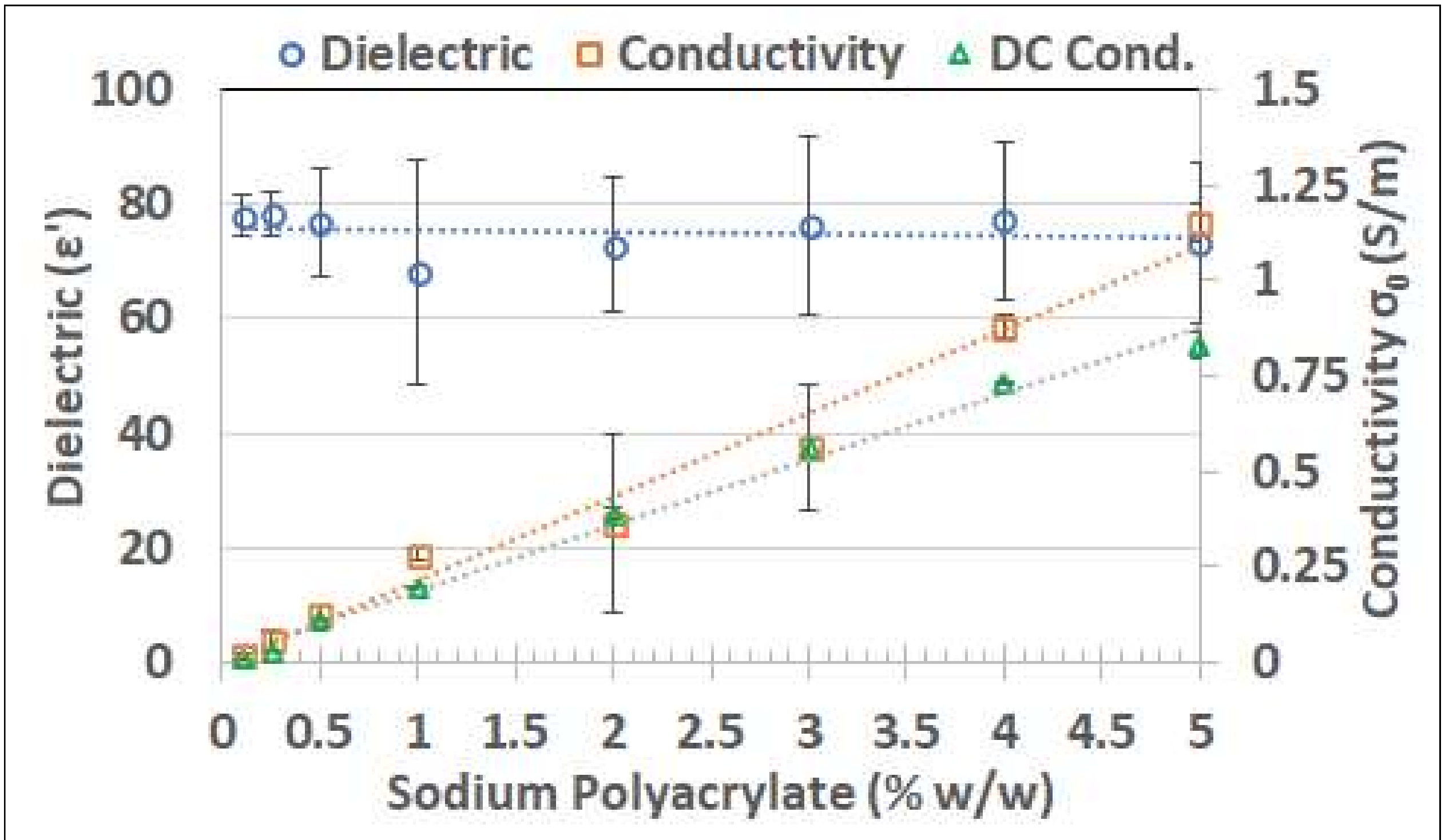

FIGURE 14 Static dielectric constant $\varepsilon_s$ (○, left axis), static electrical conductivity $\sigma_s$ (□) and DC conductivity $\sigma_{DC}$ from the conductivity meter (Δ) are shown for different concentrations of sodium polyacrylate in water. $\sigma_s$ and $\sigma_{DC}$ were highly correlated (r=0.921). The static dielectric did not vary over the tested concentrations.

TABLE 9 Sodium polyacrylate relaxation times and fit uncertainties in ms measured at room temperature 20°C. Units: $T_1$ and $T_2$ (ms), $r_1$ and $r_2$ (w/%w-s), $R_{1,0}$ and $R_{2,0}$ (1/s), *a* (w/%w), *b* (S/m·w/%w), $\sigma_{s,0}$ (S/m).

| **Concentration (%w/w)** | | **0.35 T (14.71 MHz)** | **1.5 T (63.89 MHz)** | **3 T (123.21 MHz)** |
|---|---|---|---|---|
| 0.25 | $T_1$ | 1509.67 ± 10.58 | 1663.39 ± 3.10 | 1945.18 ± 23.01 |
| | $T_2$ | 1160.79 ± 251.24 | 1068.51 ± 13.34 | 1158.53 ± 11.81 |
| 1 | $T_1$ | 1400.09 ± 18.02 | 1515.42 ± 2.92 | 1808.76 ± 8.20 |
| | $T_2$ | 1099.77 ± 52.17 | 885.06 ± 11.17 | 832.69 ± 9.98 |
| 3 | $T_1$ | 1328.95 ± 6.15 | 1446.53 ± 10.64 | 1620.51 ± 27.36 |
| | $T_2$ | 1052.61 ± 34.79 | 919.04 ± 11.52 | 765.50 ± 5.35 |
| 4 | $T_1$ | 1353.54 ± 10.31 | 1455.70 ± 4.95 | 1640.39 ± 30.76 |
| | $T_2$ | 1027.08 ± 53.71 | 868.12 ± 7.46 | 793.70 ± 8.05 |
| 5 | $T_1$ | 1364.60 ± 9.65 | 1447.16 ± 6.05 | 1670.88 ± 13.31 |
| | $T_2$ | 1022.09 ± 31.39 | 869.33 ± 7.86 | 827.81 ± 10.63 |
| Relaxivities | $r_1$ | 0.02 ± 0.01 | 0.02 ± 0.00 | 0.02 ± 0.00 |
| | $R_{1,0}$ | 0.66 ± 0.03 | 0.61 ± 0.01 | 0.53 ± 0.01 |
| | $r_2$ | 0.04 ± 0.03 | 0.11 ± 0.01 | 0.18 ± 0.01 |
| | $R_{2,0}$ | 0.93 ± 0.09 | 1.09 ± 0.04 | 1.19 ± 0.03 |

| Permittivities | | | | |
|---|---|---|---|---|
| Static Dielectric | | Static Conductivity | | DC Conductivity |
| $\varepsilon_{s,0}$ | 7.57E1 ± 1.98E0 | $\sigma_{s,0}$ | -1.76E-3 ± 3.20E-2 | 1.45E-2 ± 1.72E-2 |
| *a* | -3.27E-1 ± 7.53E-1 | *b* | 2.18E-1 ± 1.22E-2 | 1.73E-1 ± 6.55E-3 |

The $T_1$ and $T_2$ relaxation rates ($R_1$ and $R_2$, respectively) as a function of xanthan gum concentration are shown in Figures 15a and 15b The $T_1$ and $T_2$ relaxivities ($r_1$ and $r_2$, respectively) as a function of field strength are shown in Figures 15c and 15d.

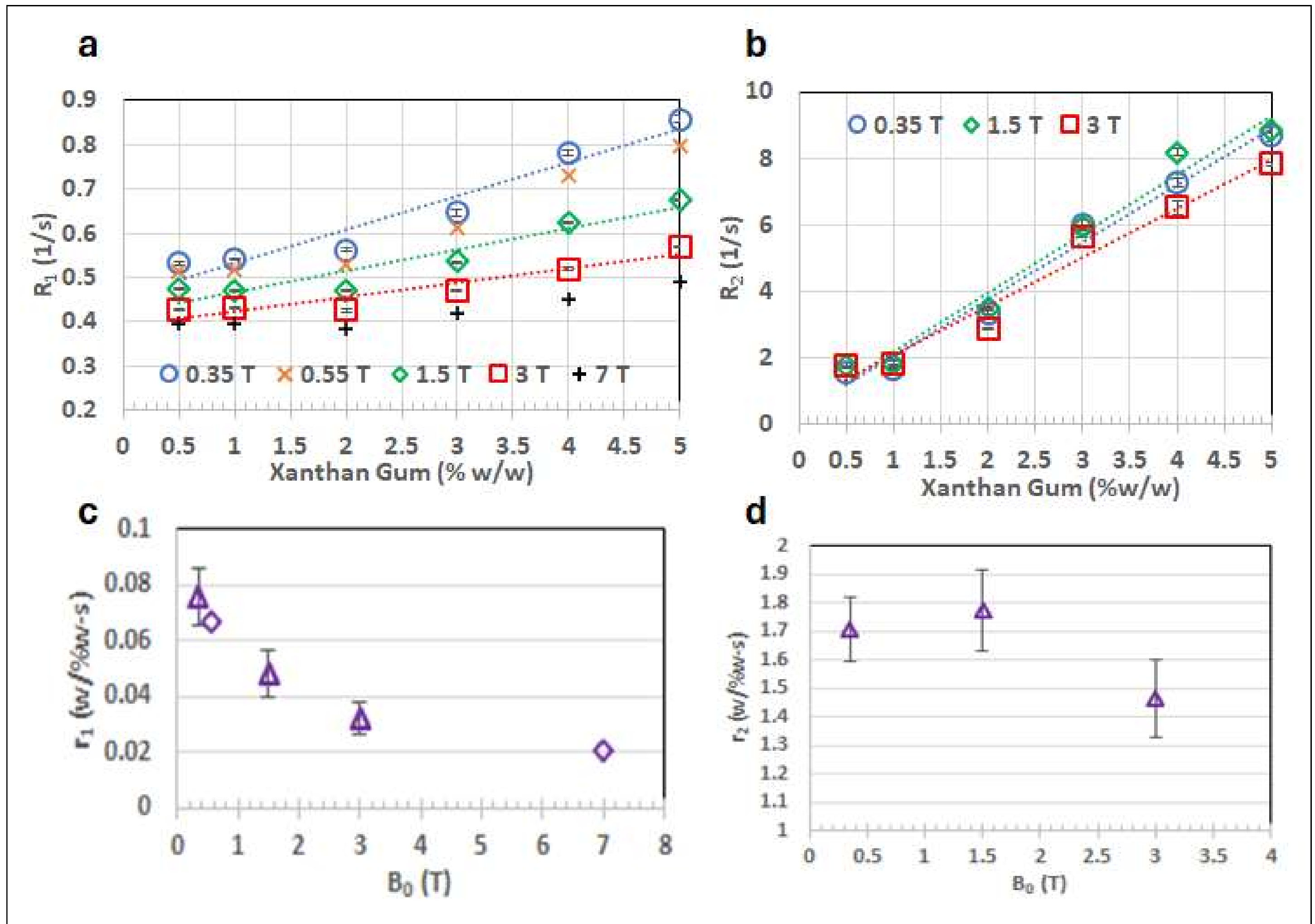

FIGURE 15 Relaxation rates for different concentrations of xanthan gum: a) $R_1$ and b) $R_2$, and xanthan gum relaxivities vs. field strength for c) $r_1$ and d) $r_2$. $R_1$ (×, +) and $r_1$ (◊) interpolations/extrapolations are included for 0.55 T and 7 T.

The dielectric strengths and electrical conductivities for different concentrations of xanthan gum are shown in Figure 16. The relaxation times, relaxivities, and permittivity measurements for xanthan gum are summarized in Table 10.

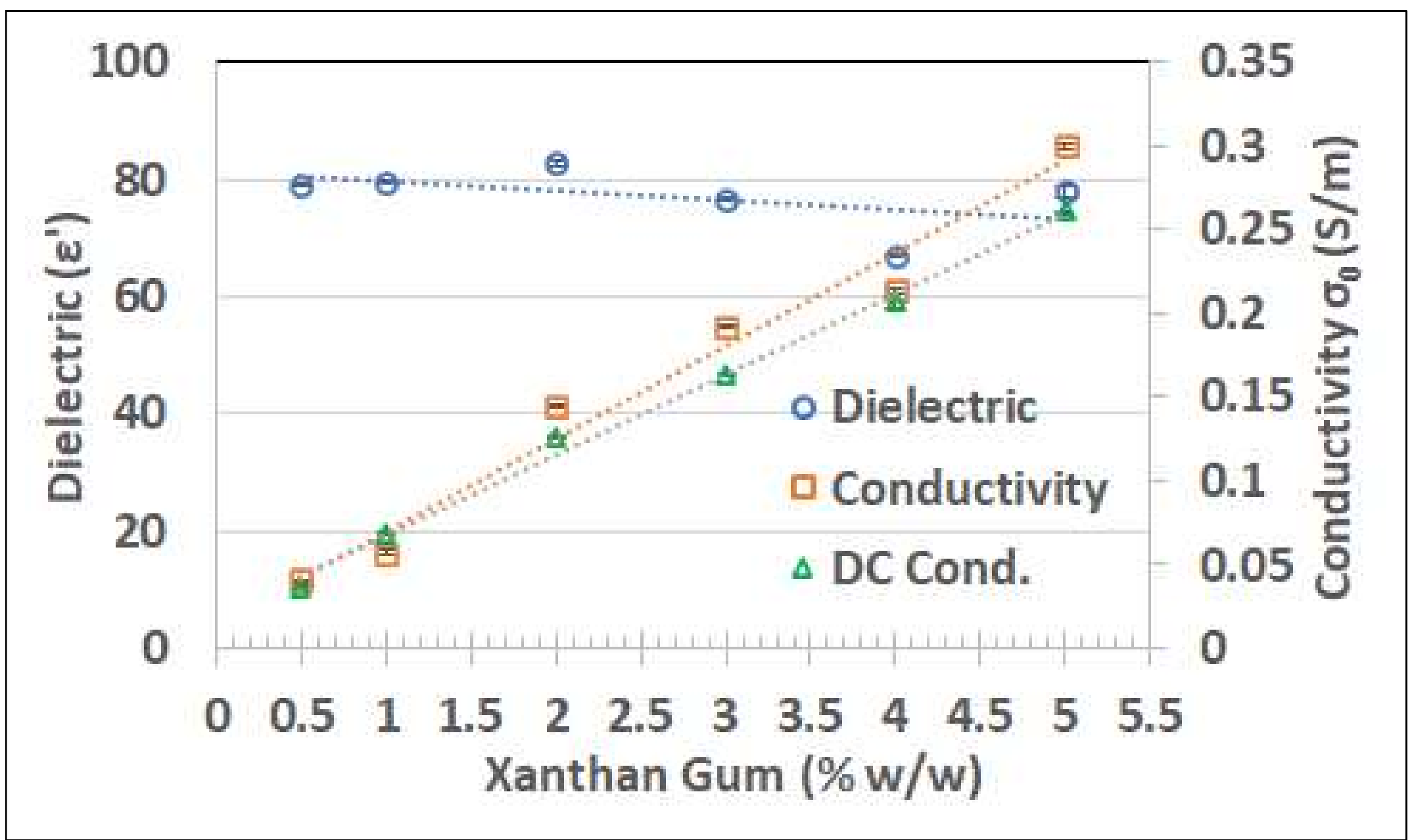

FIGURE 16 Static dielectric constant $\varepsilon_s$ (○, left axis), static electrical conductivity $\sigma_s$ (□) and DC conductivity $\sigma_{DC}$ from the conductivity meter (Δ) are shown for different concentrations of xanthan gum in water. $\sigma_s$ and $\sigma_{DC}$ were highly correlated (r=0.992).

TABLE 10 Xanthan gum relaxation times and fit uncertainties in ms measured at room temperature (20°C). Units: $T_1$ and $T_2$ (ms), $r_1$ and $r_2$ (w/%w-s), $R_{1,0}$ and $R_{2,0}$ (1/s), *a* (w/%w), *b* (S/m·w/%w), $\sigma_{s,0}$ (S/m).

| **Concentration (%w/w)** | | **0.35 T (14.71 MHz)** | **1.5 T (63.89 MHz)** | **3 T (123.21 MHz)** |
|---|---|---|---|---|
| 0.5 | $T_1$ | 1875.67 ± 12.51 | 2111.68 ± 6.05 | 2339.91 ± 11.19 |
| | $T_2$ | 653.14 ± 9.93 | 573.95 ± 3.24 | 556.93 ± 15.94 |
| 1 | $T_1$ | 1845.74 ± 7.85 | 2121.07 ± 6.70 | 2314.78 ± 10.43 |
| | $T_2$ | 596.84 ± 9.69 | 556.57 ± 5.16 | 564.19 ± 24.21 |
| 2 | $T_1$ | 1780.70 ± 12.05 | 2125.67 ± 5.22 | 2349.65 ± 15.65 |
| | $T_2$ | 297.89 ± 3.26 | 285.74 ± 3.50 | 344.01 ± 2.15 |
| 3 | $T_1$ | 1545.06 ± 19.52 | 1866.93 ± 2.67 | 2130.01 ± 11.89 |
| | $T_2$ | 166.79 ± 3.92 | 167.83 ± 1.41 | 177.94 ± 1.19 |
| 4 | $T_1$ | 1279.84 ± 8.01 | 1601.64 ± 3.14 | 1924.59 ± 14.12 |
| | $T_2$ | 137.05 ± 2.49 | 122.09 ± 1.82 | 153.27 ± 4.91 |
| 5 | $T_1$ | 1166.64 ± 11.42 | 1477.87 ± 4.01 | 1754.36 ± 8.94 |
| | $T_2$ | 115.39 ± 0.39 | 112.98 ± 1.08 | 127.75 ± 0.60 |
| Relaxivities | $r_1$ | 0.08 ± 0.01 | 0.05 ± 0.01 | 0.03 ± 0.01 |
| | $R_{1,0}$ | 0.46 ± 0.03 | 0.42 ± 0.03 | 0.39 ± 0.02 |
| | $r_2$ | 1.71 ± 0.11 | 1.77 ± 0.14 | 1.47 ± 0.14 |
| | $R_{2,0}$ | 0.34 ± 0.34 | 0.43 ± 0.44 | 0.63 ± 0.42 |

| Permittivities | | | | |
|---|---|---|---|---|
| Static Dielectric | | Static Conductivity | | DC Conductivity |
| $\varepsilon_{s,0}$ | 8.13E1 ± 4.00E0 | $\sigma_{s,0}$ | 1.45E-2 ± 1.38E-2 | 1.96E-2 ± 5.49E-3 |
| *a* | -1.57E0 ± 1.32E0 | *b* | 5.56E-2 ± 4.56E-3 | 4.82E-2 ± 1.81E-3 |

**Paramagnetic Salts:**

The $T_1$ and $T_2$ relaxation rates ($R_1$ and $R_2$, respectively) as a function of copper sulfate concentration are shown in Figures 17a and 17b The $T_1$ and $T_2$ relaxivities ($r_1$ and $r_2$, respectively) as a function of field strength are shown in Figures 17c and 17d.

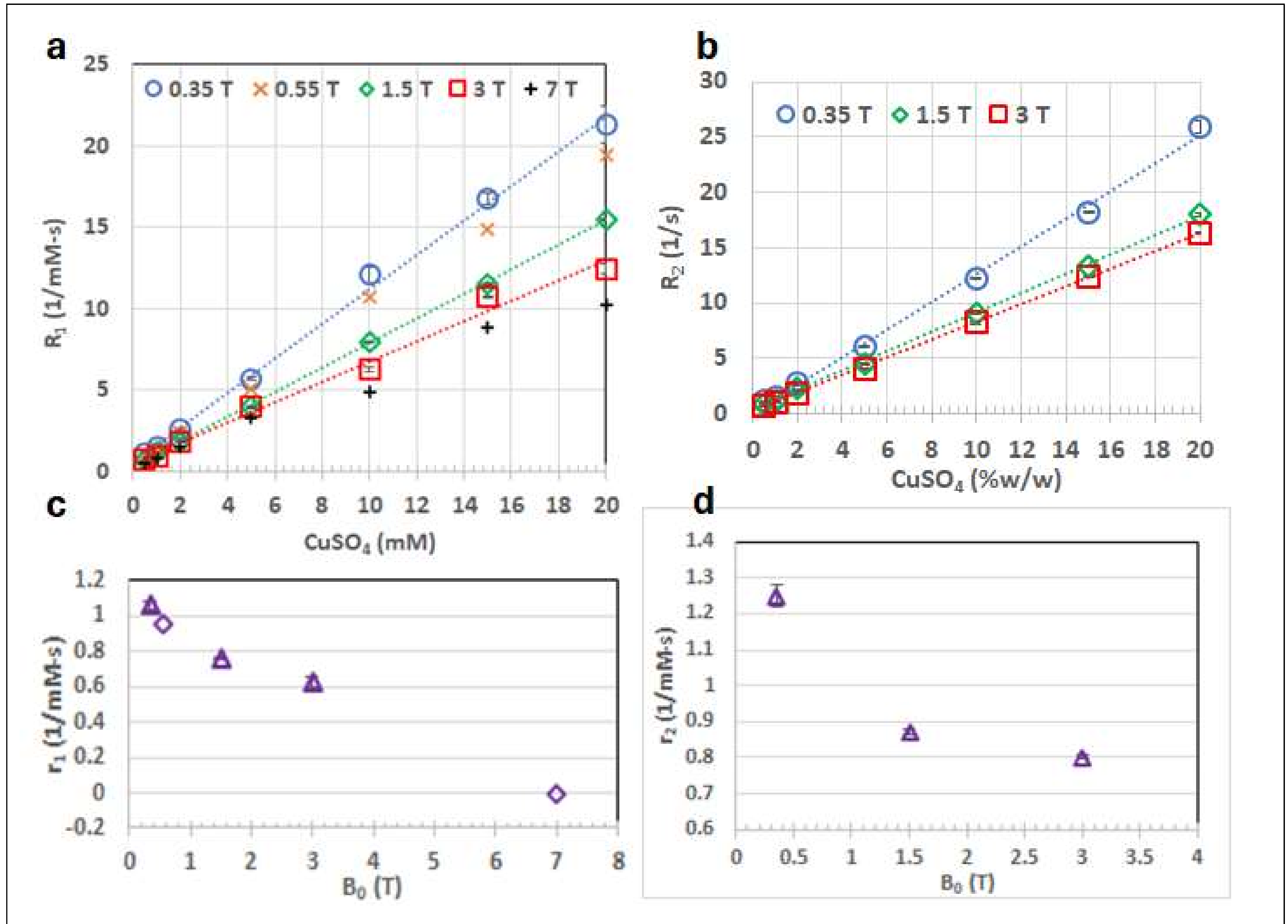

FIGURE 17 Relaxation rates for different concentrations of $CuSO_4$: a) $R_1$ and b) $R_2$, and $CuSO_4$ relaxivities vs. field strength for c) $r_1$ and d) $r_2$. $R_1$ (×, +) and $r_1$ (◊) interpolations/extrapolations are included for 0.55 T and 7 T.

The dielectric strengths and electrical conductivities for different concentrations of copper sulfate are shown in Figure 18. The relaxation times, relaxivities, and permittivity measurements for copper sulfate are summarized in Table 11.

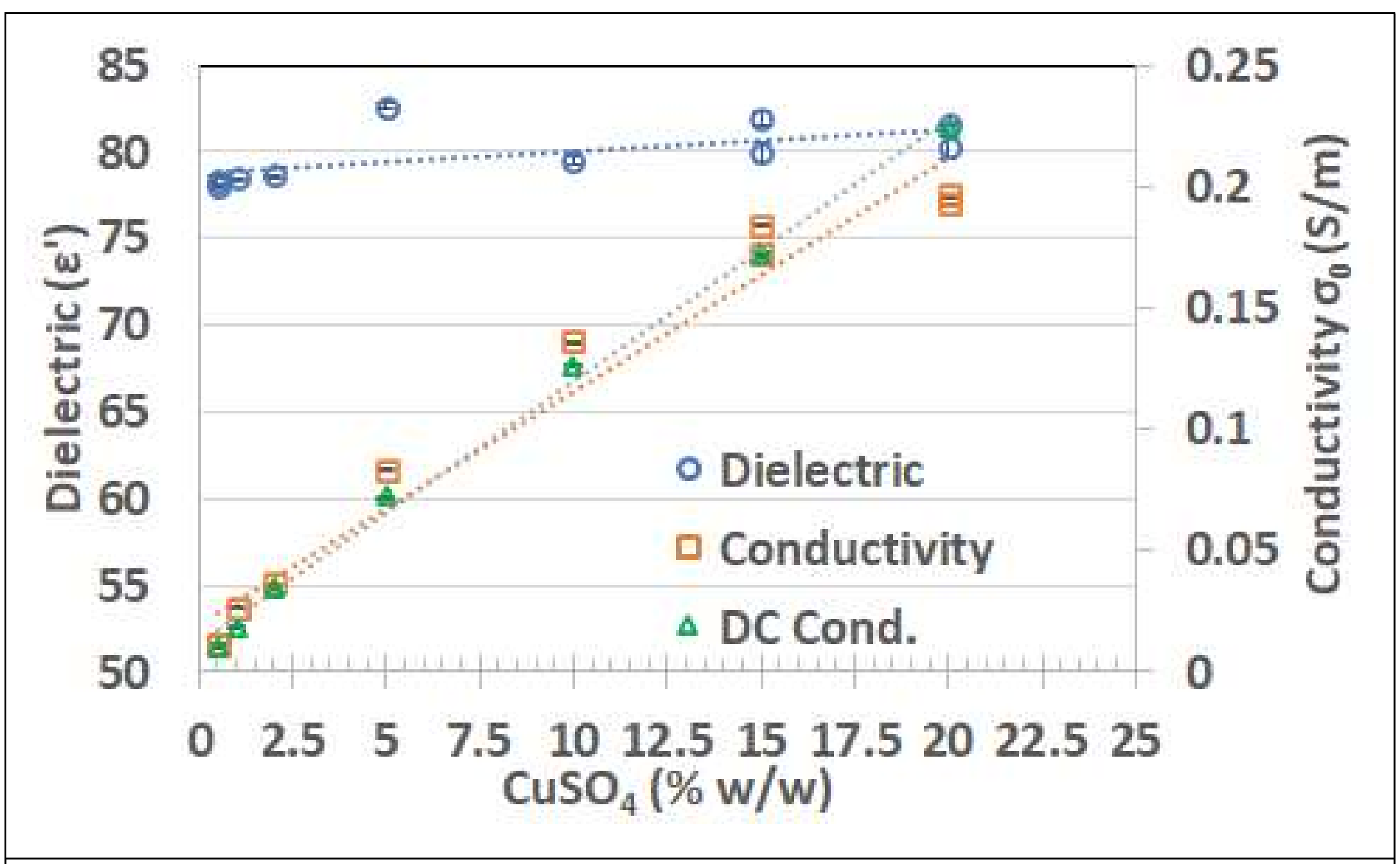

FIGURE 18 Static dielectric constant $\varepsilon_s$ (○, left axis), static electrical conductivity $\sigma_s$ (□) and DC conductivity $\sigma_{DC}$ from the conductivity meter (Δ) are shown for different concentrations of $CuSO_4$ in water. $\sigma_s$ and $\sigma_{DC}$ were highly correlated (r=0.985). $\varepsilon_s$ and $\sigma_s$ were correlated (r=-0.653).

TABLE 11 $CuSO_4$ relaxation times and fit uncertainties in ms measured at room temperature 20°C

| Concentration (mM) | | 0.35 T (14.71 MHz) | 1.5 T (63.89 MHz) | 3 T (123.21 MHz) |
|---|---|---|---|---|
| 0.5 | $T_1$ | 1058.52 ± 8.16 | 1240.05 ± 1.63 | 1361.51 ± 12.29 |
| | $T_2$ | 887.42 ± 45.11 | 1093.77 ± 10.75 | 1183.18 ± 9.82 |
| 1 | $T_1$ | 670.29 ± 10.88 | 838.55 ± 8.17 | 1035.99 ± 6.35 |
| | $T_2$ | 633.60 ± 27.37 | 804.33 ± 27.60 | 902.47 ± 49.37 |
| 2 | $T_1$ | 376.12 ± 5.83 | 536.44 ± 0.59 | 542.48 ± 10.70 |
| | $T_2$ | 365.77 ± 3.62 | 448.03 ± 12.47 | 525.49 ± 18.77 |
| 5 | $T_1$ | 174.59 ± 2.14 | 252.29 ± 3.05 | 250.99 ± 6.02 |
| | $T_2$ | 164.41 ± 1.53 | 219.51 ± 1.80 | 245.08 ± 2.60 |
| 10 | $T_1$ | 82.59 ± 3.34 | 125.04 ± 0.76 | 160.10 ± 3.34 |
| | $T_2$ | 82.15 ± 0.45 | 110.21 ± 0.50 | 121.56 ± 2.20 |
| 15 | $T_1$ | 59.65 ± 1.20 | 86.50 ± 0.25 | 92.83 ± 0.38 |
| | $T_2$ | 54.94 ± 0.45 | 74.87 ± 0.42 | 80.59 ± 0.33 |
| 20 | $T_1$ | 46.91 ± 2.53 | 64.48 ± 0.15 | 80.32 ± 1.92 |
| | $T_2$ | 38.61 ± 0.88 | 55.58 ± 0.39 | 61.35 ± 0.53 |

| Relaxivities | $r_1$ | 1.06 ± 0.03 | 0.75 ± 0.01 | 0.62 ± 0.03 |
|---|---|---|---|---|
| | $R_{1,0}$ | 0.64 ± 027 | 0.35 ± 0.03 | 0.52 ± 0.31 |
| | $r_2$ | 1.25 ± 0.03 | 0.87 ± 0.01 | 0.80 ± 0.01 |
| | $R_{2,0}$ | 0.14 ± 0.31 | 0.38 ± 0.08 | 0.28 ± 0.06 |
| Permittivities | | | | |
| Static Dielectric | | Static Conductivity | | DC Conductivity |
| $\varepsilon_{s,0}$ | 7.88E1 ± 6.67E-1 | $\sigma_{s,0}$ | 1.95E-2 ± 7.86E-3 | 1.13E-2 ± 3.15E-3 |
| *a* | 1.24E-1 ± 5.67E-2 | *b* | 9.62E-3 ± 6.69E-4 | 1.09E-2 ± 3.03E-4 |

The $T_1$ and $T_2$ relaxation rates ($R_1$ and $R_2$, respectively) as a function of manganese nitrate concentration are shown in Figures 19a and 19b The $T_1$ and $T_2$ relaxivities ($r_1$ and $r_2$, respectively) as a function of field strength are shown in Figures 19c and 19d.

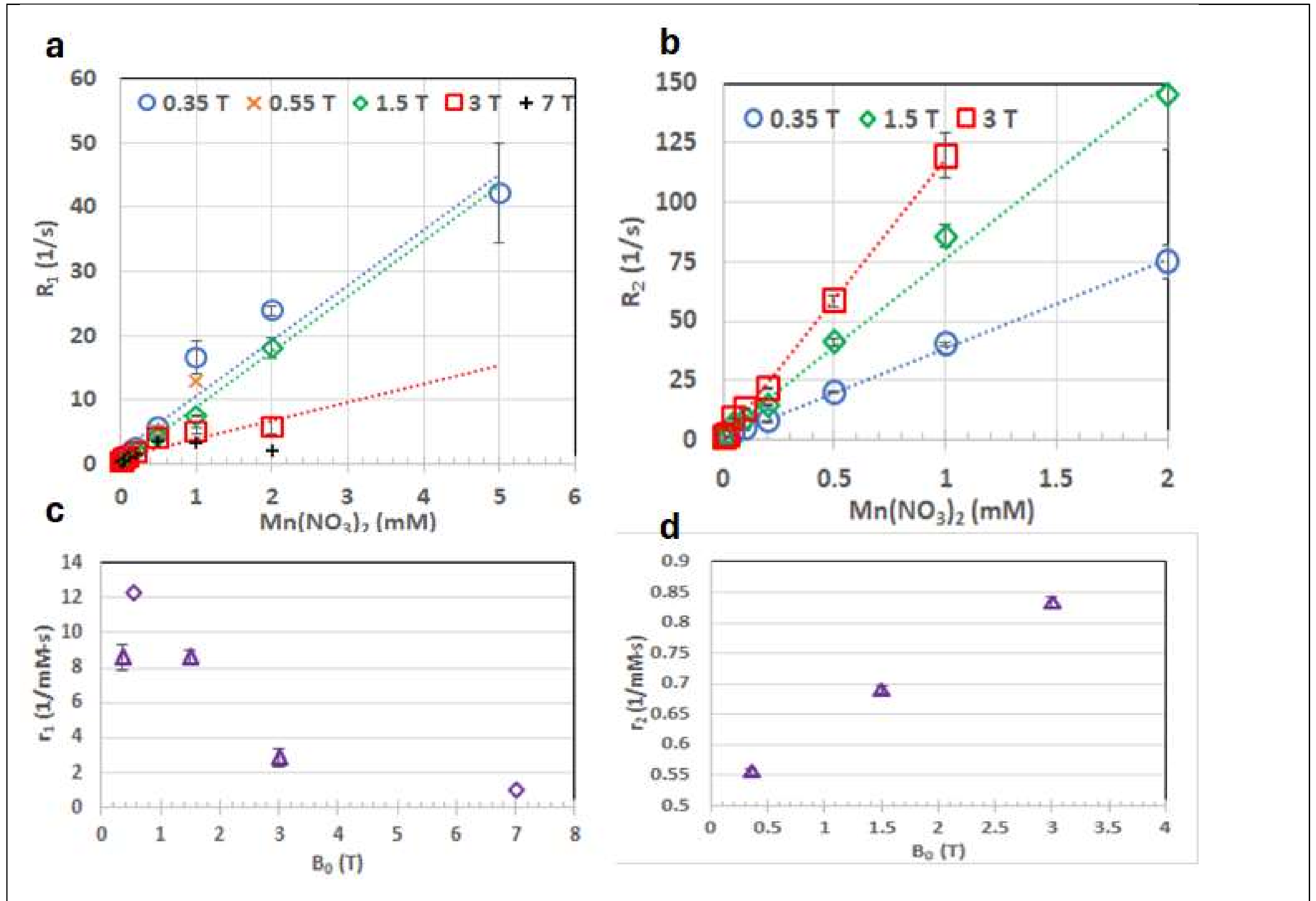

FIGURE S19 Relaxation rates for different concentrations of $Mn(NO_3)_2$: a) $R_1$ and b) $R_2$, and $Mn(NO_3)_2$ relaxivities vs. field strength for c) $r_1$ and d) $r_2$. $R_1$ (×, +) and $r_1$ (◇) interpolations/extrapolations are included for 0.55 T and 7 T. Signal dephasing was dependent on field strength and limited the range of paramagnetic salt concentrations.

The dielectric strengths and electrical conductivities for different concentrations of manganese nitrate are shown in Figure 20. The relaxation times, relaxivities, and permittivity measurements for manganese nitrate are summarized in Table 12.

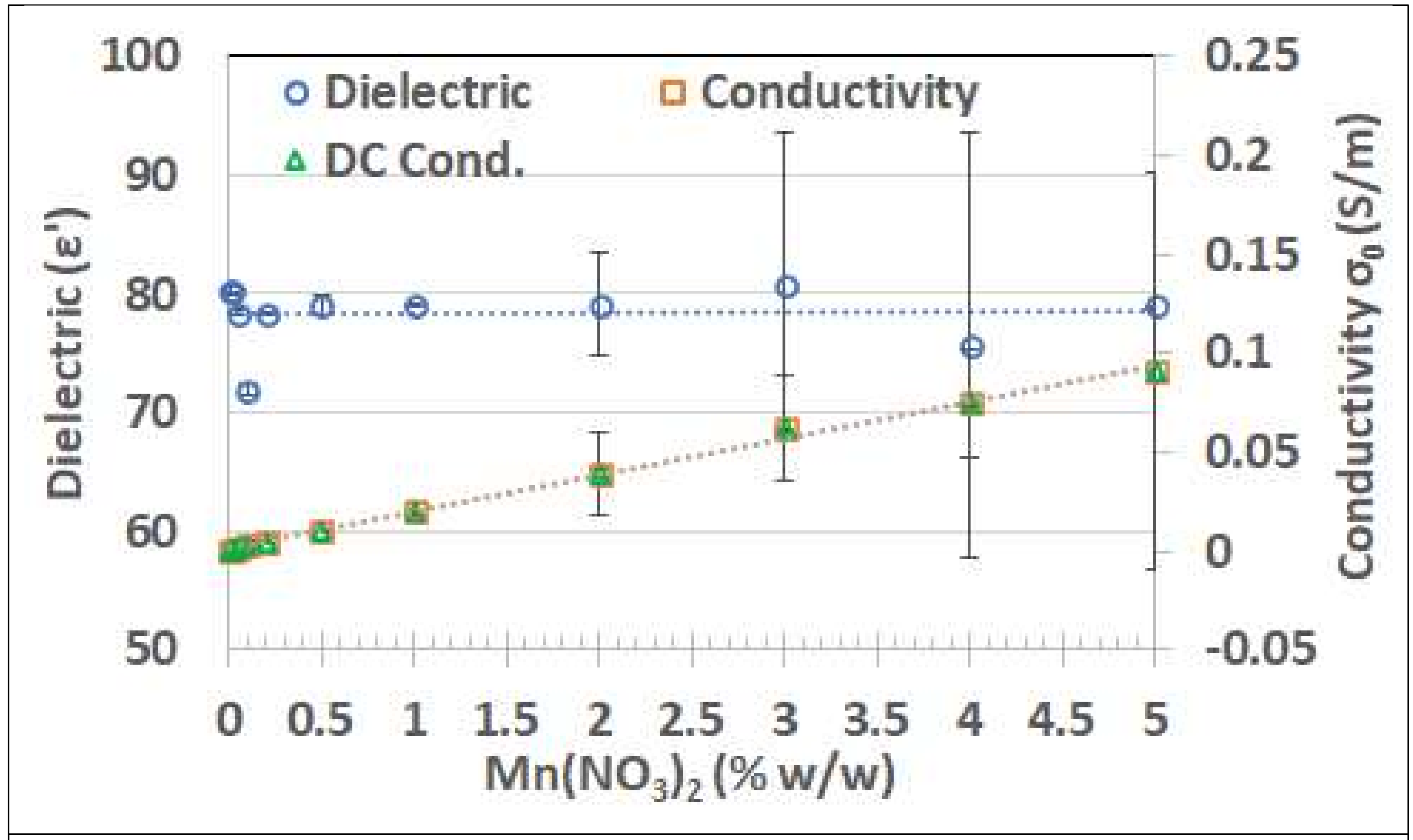

FIGURE 20 Static dielectric constant $\varepsilon_s$ (○, left axis), static electrical conductivity $\sigma_s$ (□) and DC conductivity $\sigma_{DC}$ from the conductivity meter (Δ) are shown for different concentrations of $Mn(NO_3)_2$ in water. $\sigma_s$ and $\sigma_{DC}$ were highly correlated (r=0.985).

TABLE 12 $Mn(NO_3)_2$ relaxation times and fit uncertainties in ms measured at room temperature (20°C)

| Concentration (mM) | | 0.35 T (14.71 MHz) | 1.5 T (63.89 MHz) | 3 T (123.21 MHz) |
|---|---|---|---|---|
| 0.01 | $T_1$ | 2043.11 ± 37.10 | 2202.60 ± 5.46 | 2469.91 ± 15.16 |
| | $T_2$ | 1009.49 ± 37.12 | 815.98 ± 4.41 | 672.59 ± 5.36 |
| 0.02 | $T_1$ | 1771.33 ± 6.66 | 2015.92 ± 5.21 | 2276.35 ± 34.71 |
| | $T_2$ | 966.72 ± 16.81 | 637.61 ± 3.75 | 505.15 ± 4.04 |
| 0.05 | $T_1$ | 844.51 ± 7.99 | 1061.77 ± 1.57 | 1201.41 ± 17.79 |
| | $T_2$ | 311.15 ± 5.51 | 152.56 ± 2.34 | 112.75 ± 5.17 |
| 0.1 | $T_1$ | 574.24 ± 3.77 | 746.85 ± 4.62 | 815.71 ± 3.41 |
| | $T_2$ | 198.02 ± 2.68 | 109.99 ± 0.67 | 77.19 ± 0.89 |
| 0.2 | $T_1$ | 369.90 ± 4.05 | 464.32 ± 0.91 | 548.32 ± 12.29 |
| | $T_2$ | 123.71 ± 0.85 | 67.06 ± 0.41 | 45.92 ± 0.69 |
| 0.5 | $T_1$ | 171.62 ± 1.68 | 226.82 ± 0.81 | 242.84 ± 1.50 |
| | $T_2$ | 49.34 ± 0.74 | 24.25 ± 0.82 | 17.09 ± 0.67 |
| 1 | $T_1$ | 87.31 ± 1.40 | 118.73 ± 1.45 | 224.37 ± 19.08 |
| | $T_2$ | 24.53 ± 1.09 | 11.04 ± 0.87 | 8.37 ± 0.67 |
| 2 | $T_1$ | 41.82 ± 1.32 | 55.05 ± 4.87 | 171.85 ± 36.43 |
| | $T_2$ | 13.32 ± 1.22 | 6.90 ± 1.09 | * |
| 3 | $T_1$ | 10.94 ± 1.34 | 46.81 ± 3.35 | * |
| | $T_2$ | 7.42 ± 0.42 | 4.12 ± 0.46 | * |
| Relaxivities | $r_1$ | 8.60 ± 0.71 | 8.62 ± 0.34 | 2.86 ± 0.55 |
| | $R_{1,0}$ | 2.10 ± 1.30 | 0.26 ± 0.28 | 1.09 ± 0.45 |
| | $r_2$ | 37.46 ± 0.50 | 73.80 ± 2.55 | 118.35 ± 1.91 |
| | $R_{2,0}$ | 1.08 ± 0.41 | 2.36 ± 2.08 | 0.36 ± 0.82 |

| Permittivities | | | | |
|---|---|---|---|---|
| Static Dielectric | | Static Conductivity | | DC Conductivity |
| $\varepsilon_{s,0}$ | 7.83E1 ± 1.04E0 | $\sigma_{s,0}$ | 4.19E-3 ± 2.07E-3 | 1.19E-3 ± 7.83E-4 |
| *a* | 5.10E-2 ± 4.65E-1 | *b* | 1.55E-2 ± 9.21E-4 | 1.86E-2 ± 3.49E-4 |

*Significant signal dephasing prevents quantification.

The $T_1$ and $T_2$ relaxation rates ($R_1$ and $R_2$, respectively) as a function of nickel chloride concentration are shown in Figures 21a and 21b The $T_1$ and $T_2$ relaxivities ($r_1$ and $r_2$, respectively) as a function of field strength are shown in Figures 21c and 21d.

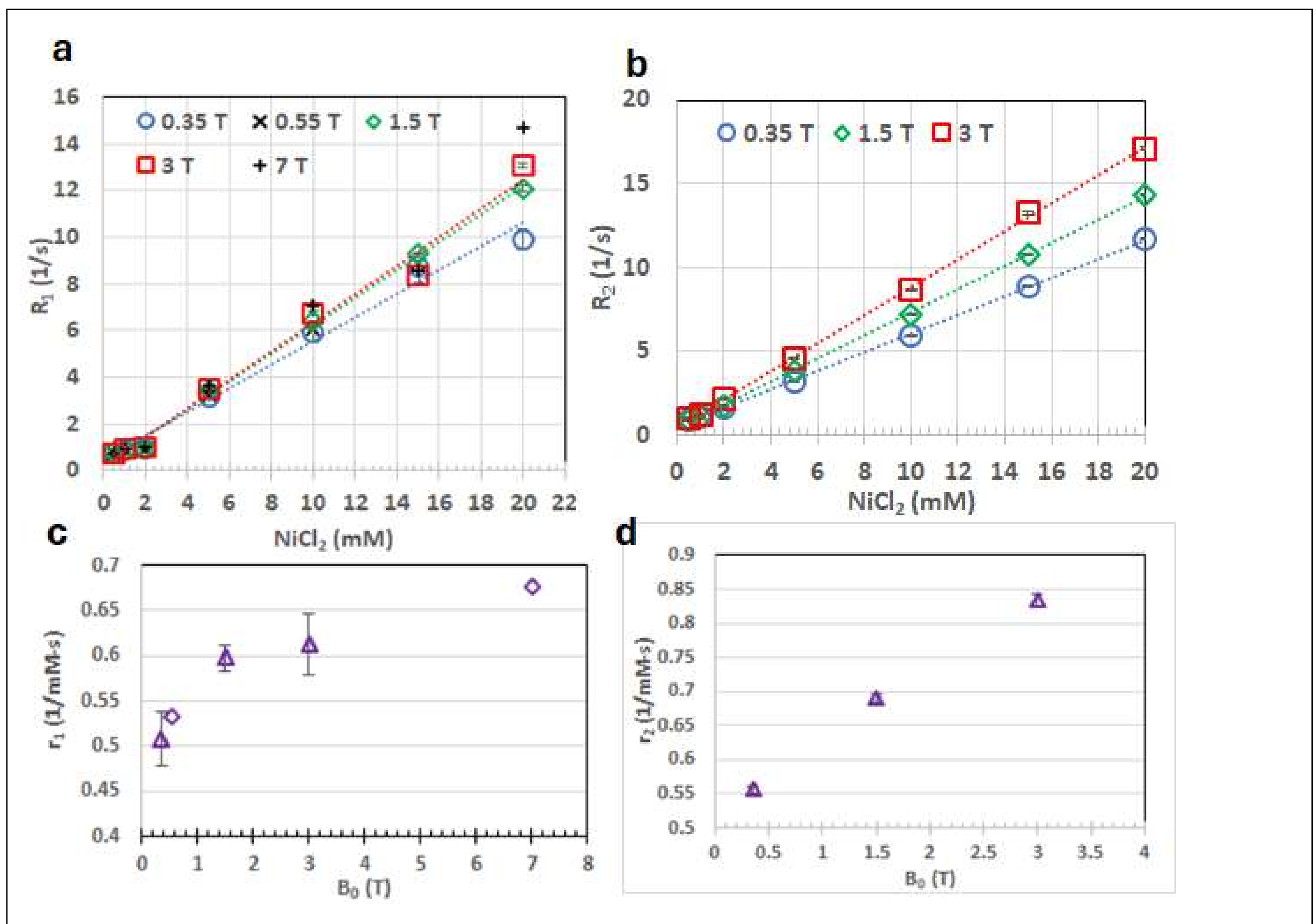

FIGURE 21 Relaxation rates for different concentrations of $NiCl_2$: a) $R_1$ and b) $R_2$, and $NiCl_2$ relaxivities vs. field strength for c) $r_1$ and d) $r_2$. $R_1$ (×, +) and $r_1$ (◊) interpolations/extrapolations are included for 0.55 T and 7 T.

The dielectric strengths and electrical conductivities for different concentrations of nickel chloride are shown in Figure 22. The relaxation times, relaxivities, and permittivity measurements for nickel chloride are summarized in Table 13.

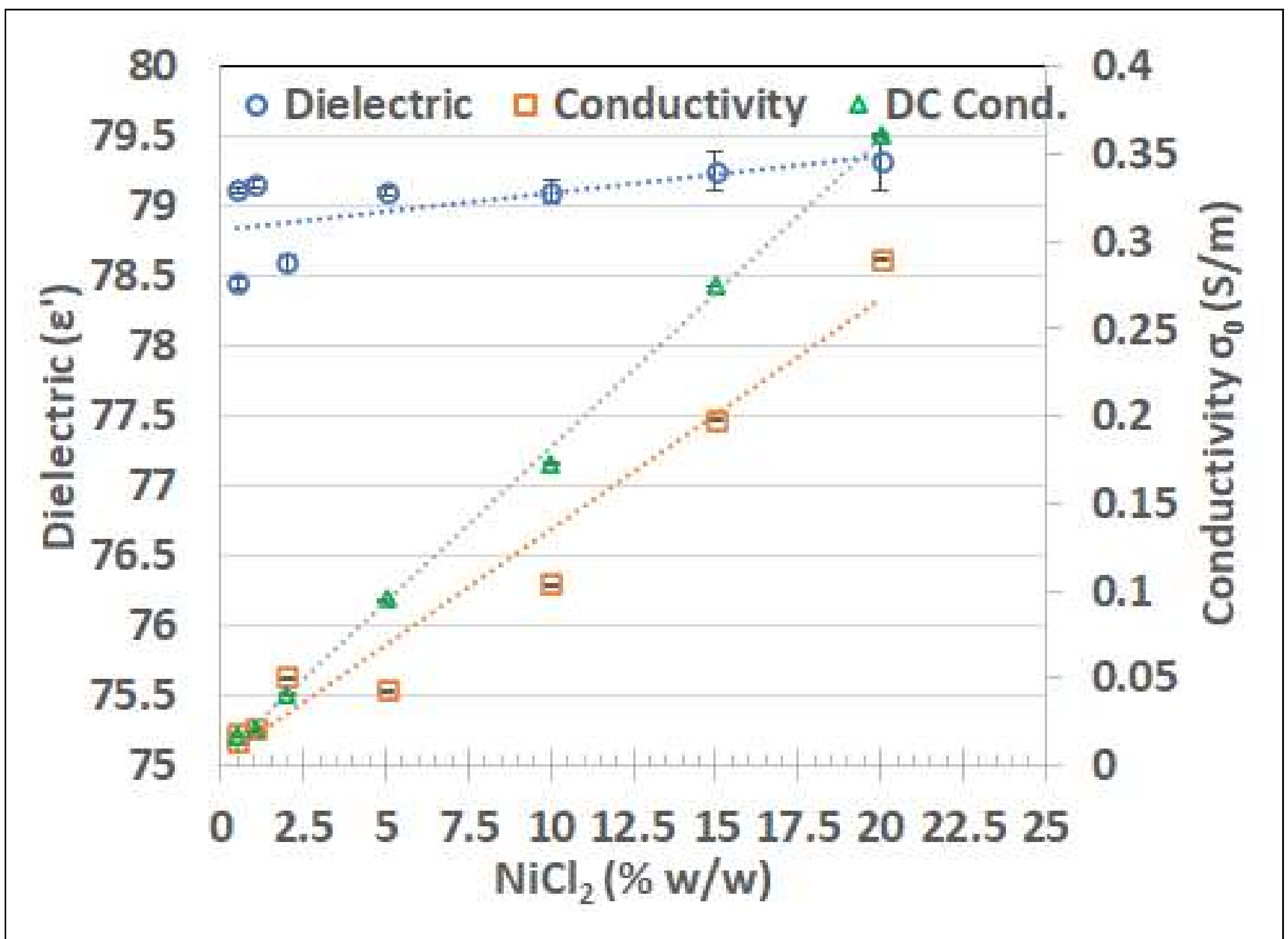

FIGURE 22 Static dielectric constant $\varepsilon_s$ (○, left axis), static electrical conductivity $\sigma_s$ (□) and DC conductivity $\sigma_{DC}$ from the conductivity meter (Δ) are shown for different concentrations of $NiCl_2$ in water. $\sigma_s$ and $\sigma_{DC}$ were highly correlated (r=0.981) but significantly different. $\varepsilon_s$ and $\sigma_s$ were correlated (r=-0.653).

TABLE 13 $NiCl_2$ relaxation times and fit uncertainties in ms measured at room temperature (20°C)

| Concentration (mM) | | 0.35 T (14.71 MHz) | 1.5 T (63.89 MHz) | 3 T (123.21 MHz) |
|---|---|---|---|---|
| 0.5 | $T_1$ | 1341.56 ± 31.87 | 1293.22 ± 2.89 | 1324.84 ± 12.29 |
| | $T_2$ | 1084.03 ± 14.17 | 1058.52 ± 12.69 | 969.98 ± 11.06 |
| 1 | $T_1$ | 1052.18 ± 8.05 | 1031.44 ± 2.57 | 1023.46 ± 4.22 |
| | $T_2$ | 927.20 ± 9.69 | 860.28 ± 4.24 | 792.25 ± 3.84 |
| 2 | $T_1$ | 543.86 ± 3.47 | 513.58 ± 3.04 | 502.96 ± 19.60 |
| | $T_2$ | 504.28 ± 5.25 | 447.17 ± 12.47 | 381.18 ± 0.49 |
| 5 | $T_1$ | 314.77 ± 3.14 | 294.12 ± 3.24 | 288.09 ± 3.78 |
| | $T_2$ | 307.58 ± 4.89 | 262.23 ± 1.41 | 217.20 ± 1.12 |
| 10 | $T_1$ | 168.16 ± 12.34 | 155.57 ± 0.58 | 148.49 ± 3.78 |
| | $T_2$ | 167.11 ± 0.67 | 137.91 ± 0.59 | 115.52 ± 0.55 |
| 15 | $T_1$ | 113.91 ± 2.61 | 107.37 ± 0.13 | 120.17 ± 4.04 |
| | $T_2$ | 112.24 ± 0.60 | 92.97 ± 0.30 | 75.34 ± 0.63 |
| 20 | $T_1$ | 101.07 ± 4.40 | 82.82 ± 0.95 | 76.51 ± 4.04 |
| | $T_2$ | 85.49 ± 0.46 | 69.68 ± 0.29 | 58.53 ± 0.41 |
| Relaxivities | $r_1$ | 0.51 ± 0.03 | 0.60 ± 0.01 | 0.61 ± 0.03 |
| | $R_{1,0}$ | 0.45 ± 0.31 | 0.26 ± 0.15 | 0.20 ± 0.35 |
| | $r_2$ | 0.56 ± 0.00 | 0.69 ± 0.01 | 0.83 ± 0.01 |
| | $R_{2,0}$ | 0.54 ± 0.04 | 0.44 ± 0.06 | 0.49 ± 0.09 |

| Permittivities | | | | |
|---|---|---|---|---|
| Static Dielectric | | Static Conductivity | | DC Conductivity |
| $\varepsilon_{s,0}$ | 7.88E1 ± 1.32E-1 | $\sigma_{s,0}$ | 3.09E-3 ± 1.05E-2 | 5.02E-3 ± 2.67E-3 |
| *a* | 2.61E-2 ± 1.36E-2 | *b* | 1.32E-2 ± 1.08E-3 | 1.77E-2 ± 2.57E-4 |

**Oils:**

The $T_1$ and $T_2$ relaxation rates ($R_1$ and $R_2$, respectively) as a function of field strength are shown in Figures 23a and 23b for canola, castor, and grapeseed oils. Deionized water and 0.9% (normal) saline are shown for comparison. The relaxation times are documented in Table 14. The permittivity results are shown in Table 15.

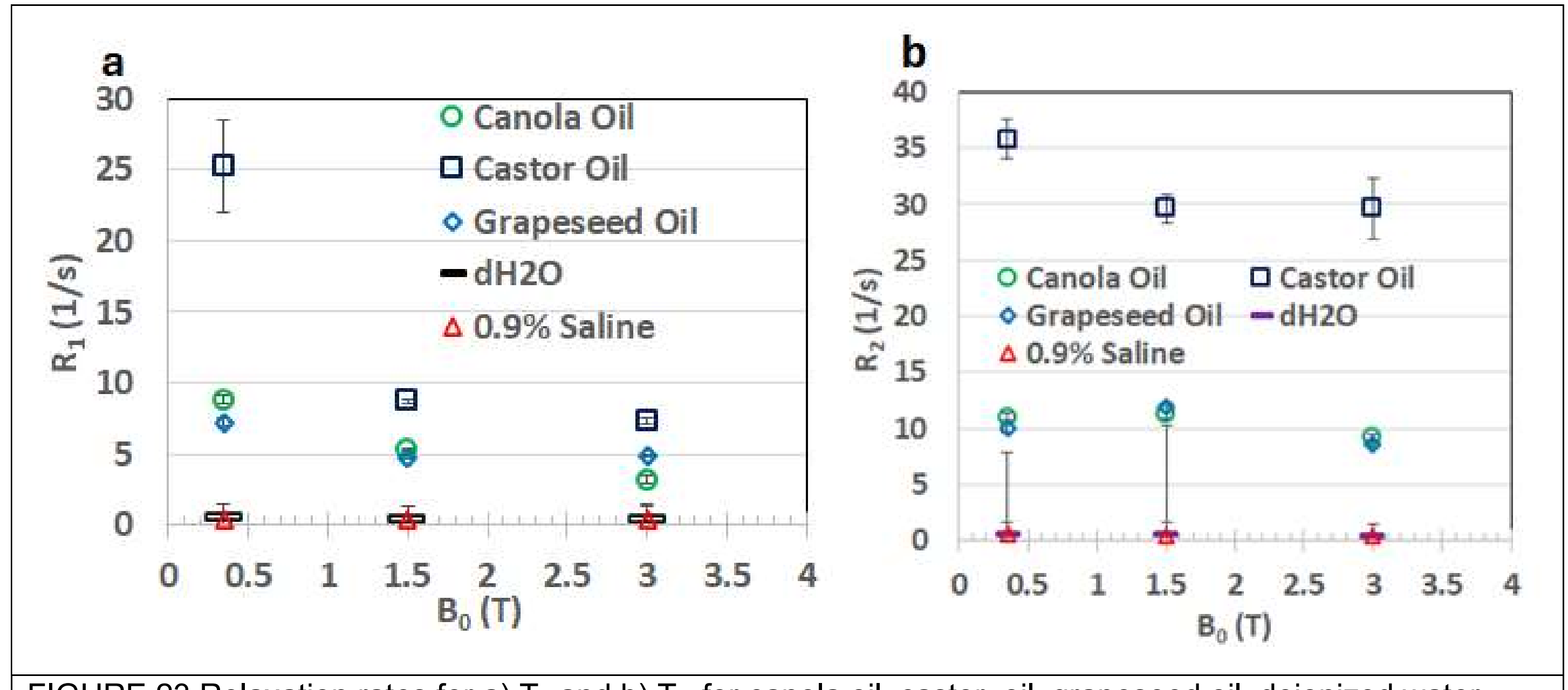

FIGURE 23 Relaxation rates for a) $T_1$ and b) $T_2$. for canola oil, castor, oil, grapeseed oil, deionized water, and 0.9% saline.(dH2O),

TABLE 14 Liquid relaxation times and fit uncertainties in ms measured at room temperature (20°C)

| **Sample** | | **0.35 T (14.71 MHz)** | **1.5 T (63.89 MHz)** | **3 T (123.21 MHz)** |
|---|---|---|---|---|
| Canola oil | $T_1$ | 126.87 ± 4.46 | 182.24 ± 2.98 | 214.11 ± 18.90 |
| | $T_2$ | 97.51 ± 1.87 | 106.09 ± 2.21 | 127.72 ± 1.29 |
| Castor oil | $T_1$ | 44.10 ± 8.05 | 123.56 ± 2.70 | 143.44 ± 16.46 |
| | $T_2$ | 24.75 ± 0.75 | 33.84 ± 0.70 | 40.15 ± 0.79 |
| Grapeseed oil | $T_1$ | 133.67 ± 9.58 | 215.21 ± 4.59 | 258.88 ± 1.63 |
| | $T_2$ | 85.86 ± 3.00 | 101.97 ± 1.83 | 116.37 ± 6.76 |

| dH$_2$O | $T_1$ | 2184.01 ± 46.67 | 2563.73 ± 6.22 | 2522..47 ± 40.20 |
| --- | --- | --- | --- | --- |
| | $T_2$ | 2051.72 ± 12.84 | 1755.22 ± 13.42 | 2422.19 ± 16.67 |
| 0.9% Saline | $T_1$ | 2466.16 ± 17.57 | 2622.79 ± 18.14 | 2791.69 ± 12.57 |
| | $T_2$ | 1786.26 ± 130.187 | 2458.1 ± 112.11 | 2724.71 ± 259.18 |

TABLE 15 Permittivities and conductivities measured at room temperature (20°C)

| **Sample** | **$\varepsilon_s$** | **$\sigma_s$ (S/m)** |
| --- | --- | --- |
| Canola oil | 4.96E0 ± 1.69E0 | 6.87E-4 ± 4.30E-5 |
| Castor oil | 3.16E0 ± 1.27E-2 | 1.98E-3 ± 1.03E-4 |
| Grapeseed oil | 3.34E0 ± 1.5E-3 | 5.00E-3 ± 5.44E-4 |
| dH$_2$O | 7.96E1 ± 1.06E-3 | 1.54E-4 ± 4.38E-5 |
| 0.9% Saline | 7.74E1 ± 5.80E-3 | 1.58E0 ± 6.50E-4 |

The viscosities measured at $22^0$C were: canola oil (70.54 +/- 0.09 mm$^2$/s), castor oil (826.62 +/- 7.55 mm$^2$/s),and grapeseed oil (61.42 +/- 0.25 mm$^2$/s).[27] Proton densities relative to water were 0.96 (castor oil), 1.02 (canola oil), and 0.93 (grapeseed oil).

## 4. CONCLUSIONS

In principle, $T_1$ of tissue should increase with field strength while $T_2$ should not vary with field strength.[28,29] However, the $T_1$ of $NiCl_2$ did not rise with field strength. We observed $T_2$ decreasing with increasing field strength in PEG, PVA, sodium polyacrylate, and $Mn(NO_3)_2$. $T_2$ increased with field strength in $CuSO_4$.

$T_1$ relaxivities ($r_1$) may drop with rising field strength.[30] However, $r_1$ remained constant in PEG and PVP, and rose for $NiCl_2$. $T_2$ relaxivities ($r_2$) typically remain constant or rise with rising field strength. Yet $r_2$ dropped for $CuSO_4$. The 0.35 T relaxivities for $Cu^{+2}$ and $Ni^{+2}$ were similar to relaxivities at 0.5 T.[31]

Phantoms for measuring RF heating require electrical conductivities similar to *in vivo* (0-1.5 S/m). Saline's dielectric decreases and conductivity increases with salinity.[32] Addition of paramagnetic electrolytes also increases electrical conductivities. $Mn(NO_3)_2$ had significantly higher relaxation rates compared to $CuSO_4$ and $NiCl_2$. Our 3 T relaxivity results were similar to those from a 3 T study.[15] Our 0.35 T relaxation times and relaxivities for $Ni^{+2}$ and $Cu^{+2}$ were similar to 0.5 T values from an earlier study.[31] Caution must be used with the paramagnetic electrolytes since they can be corrosive ($Cu^{+2}$, $Mn^{+2}$, and $Ni^{+2}$), neurotoxic ($Mn^{+2}$), and carcinogenic ($Cu^{+2}$ and $Ni^{+2}$). Our 1.5 T $CuSO_4$ relaxation times were consistent with an earlier study.[33]

QA for high field strengths ($B_0$>1.5 T) requires low dielectric phantoms to avoid dielectric artifacts. Oils typically have low dielectrics and electrical conductivities.[34] $\varepsilon_s$ was slowly varying for hydrogel concentration except for PVP which can drop $\varepsilon_s$ to <40.[35,36] PVP also has a sharp dominant peak that can minimize chemical shift artifacts.

Multiple compartment phantom materials were modeled to adjust $T_1$ and $T_2$ to be similar to specific tissues.[7 37-40] Conductive oils that combine polymers (e.g., polypropylene glycol) and metal (e.g., zinc oxide) may present an opportunity.

Preparation of homogeneous hydrogels was a challenge in this study. Originally, the Nambu freezing and thawing method was used to prepare the PVA gel at concentrations ranging from 10% to 30% (w/w).[5] However, the resulting gel was too firm for the purpose of this experiment, so the simple aqueous solution described was prepared at lower concentrations instead. Nevertheless, the heterogeneity of the hydrogels often resulted in high variance in the relaxation rates. Combinations of materials have been published.[17,41] Longevities of <3 months are typically assumed.[25]

Synthetic polymers are more desirable in phantoms than organic polymers since the synthetic polymers are more resistant to organic growth.[42] The paramagnetic electrolytes tend to prevent microorganisms from growing in the phantoms. They also increase the salinity and conductivity of the resulting liquid. However, electrolytes can impact gelling so higher concentrations of the gelling agent may be necessary.[43]

PVP is used in diffusion phantom manufacturing.[8] Our $T_2$ values for PVP were significantly higher than other studies.[44] Our sodium alginate results were similar to relaxation times measured at 2.35 T.[45] Our measured relaxivities for PEG were small for the tested range.

In summary, sodium polyacrylate, PVP, and PEG hydrogels were challenging for generating homogeneous phantoms. Sodium alginate, Miller's LB agar, and xanthan gum had good properties although they are vulnerable to biological growth. Generally, hydrogels have a limited life while paramagnetic electrolytes and oils may have a long life.

All three oils (canola, castor, and grapeseed) were good candidates for low dielectric ($\varepsilon_s$<10) phantoms (e.g., high $B_0$) with low conductivities. $Mn(NO_3)_2$ had the highest relaxivities of the paramagnetic electrolytes tested.

Our study is unique since it covers variations of phantom materials versus concentration and field strength. The main limitation is we characterized single (oil) or duplex (water plus gel or electrolyte) materials.